\documentclass[a4paper,11pt]{article}
\pdfoutput=1 
\RequirePackage{snapshot}  

\usepackage{jheppub} 

\usepackage{bm,amsmath,amssymb,slashed,graphicx,%
            enumerate,alltt,xspace,multirow,xcolor,mathrsfs}
\usepackage{fancyvrb}
\usepackage{booktabs,tabularx}
\usepackage{graphicx}
\usepackage{subcaption}
\usepackage{xspace}
\usepackage{cancel}
\usepackage[utf8]{inputenc} 
\usepackage[compat=1.0.0]{tikz-feynman}
\usepackage{scalerel}
\usepackage{layouts}
\usepackage{adjustbox}
\usepackage{pdflscape}
\usepackage[normalem]{ulem} 

\usepackage{printlen}
\usepackage{wasysym}

\newcommand{\ps}{\text{\sc ps}}
\newcommand{\heg}{\text{\sc heg}}

\newcommand\Rps{R_\text{\sc ps}}

\newcommand\BB{\bar{B}}

\newcommand{\muR}{\mu_\text{\textsc{r}}}
\newcommand{\xR}{x_\text{\textsc{r}}}
\newcommand{\xhard}{x_\text{hard}}

\newcommand\betaps{\beta_\text{\textsc{ps}}}
\newcommand\betaobs{\beta_\text{obs}}

\newcommand{\mathd}{\mathrm{d}}

\newcommand{\PhiB}{\Phi_{\rm B}}
\newcommand{\PhiBR}{\Phi}
\newcommand{\Phirad}{\Phi_{\rm rad}}

\newcommand\pythiaeight{{Pythia$\,$8}\xspace}
\newcommand\pspythiaeight{{PSPythia$\,$8}\xspace}

\newcommand\MCatNLO{MC@NLO\xspace}
\newcommand\MAcNLOPS{MAcNLOPS\xspace}
\newcommand\POWHEG{POWHEG\xspace}
\newcommand\powheg{\POWHEG}
\newcommand\powhegbeta{\POWHEG{$_\beta$}\xspace}
\newcommand\POWHEGbeta{\POWHEG{$_\beta$}\xspace}

\usepackage[capitalise]{cleveref}
\makeatletter
\g@addto@macro\bfseries{\boldmath}
\makeatother

\definecolor{labelkey}{rgb}{0,0.5,0.0}
\definecolor{royalpurple}{rgb}{0.47, 0.32, 0.66}

\usepackage{listings}
\definecolor{darkgreen}{rgb}{0,0.4,0}
\definecolor{grey}{rgb}{0.5,0.5,0.5}
\definecolor{rust}{rgb}{0.9,0.4,0.0}

\allowdisplaybreaks 

\definecolor{semiblue}{rgb}{0.3,0.3,0.8}
\newcommand{\logbook}[2]{}

\newcommand{\GeV}{\;\mathrm{GeV}}
\newcommand{\TeV}{\;\mathrm{TeV}}
\newcommand{\order}[1]{\mathcal{O}\left(#1\right)}
\newcommand{\as}{\alpha_s}

\newcommand{\ptilde}{{\widetilde p}}

\newcommand{\abar}{{\bar \alpha}}

\newcommand{\itilde}{{\tilde \imath}}
\newcommand{\jtilde}{{\tilde \jmath}}

\tikzstyle{block} = [rectangle, minimum width=1.0cm, minimum height=0.75cm, thin, draw=black]
\tikzstyle{blob} = [circle, minimum width=0.5cm, thin, draw=black]
\tikzset{blackarrow/.style={-stealth, semithick, draw=black}}
\tikzset{connection/.style={inner sep=0,outer sep=0}}

\newcolumntype{C}{>{\centering\arraybackslash}X}

\title{Matching and event-shape NNDL accuracy in parton showers}

\preprint{OUTP-23-01P, CERN-TH-2023-004}

\newcommand{\OXaff}{Rudolf Peierls Centre for Theoretical Physics, Clarendon Laboratory, Parks Road,
  University of Oxford, Oxford OX1 3PU, UK}
\newcommand{\ASCaff}{All Souls College, Oxford OX1 4AL, UK}
\newcommand{\CERNaff}{CERN, Theoretical Physics Department, CH-1211 Geneva 23, Switzerland}

\author[a]{Keith~Hamilton,}%
\author[b,c]{Alexander Karlberg,}%
\author[b,d]{Gavin~P.~Salam,}%
\author[b]{Ludovic Scyboz,}%
\author[a]{Rob Verheyen}%

\affiliation[a]{Department of Physics and Astronomy, University College London, London, WC1E 6BT, UK}
\affiliation[b]{\OXaff}
\affiliation[c]{\CERNaff}
\affiliation[d]{\ASCaff}

\date{Received: date / Accepted: \today}

\abstract{
  To explore the interplay of NLO matching and next-to-leading
  logarithmic (NLL) parton showers, we consider the simplest case of
  $\gamma^*$ and Higgs-boson decays to $q\bar q$ and $gg$
  respectively.
  Not only should shower NLL accuracy be retained across observables
  after matching, but for global event-shape observables and the
  two-jet rate, matching can augment the shower in such a way that it
  additionally achieves next-to-next-to-double-logarithmic (NNDL) accuracy, a first
  step on the route towards general NNLL.
  As a proof-of-concept exploration of this question, we consider
  direct application of multiplicative matrix-element corrections, as
  well as simple implementations of \MCatNLO and \POWHEG-style
  matching.
  We find that the first two straightforwardly bring NNDL accuracy,
  and that this can also be achieved with \POWHEG, although particular
  care is needed in the handover between \POWHEG and the shower.
  Our study involves both analytic and
  numerical components and we also touch on some
  phenomenological considerations.
}

\keywords{QCD, Parton Shower, NLO, Matching, Resummation, LHC, LEP
  \\[4em]
  \textit{For the purpose of Open Access, the authors have applied a CC BY
  public copyright licence to any Author Accepted Manuscript (AAM)
  version arising from this submission.}
}

\begin{document}

\maketitle

\section{Introduction}

\label{sec:intro}
Next-to-leading order (NLO) accurate event generators have become the
{\it de facto} tool for the simulation of particle collisions at the
LHC.
This is in large part due to the success of the two most
widely-used NLO matching schemes, \MCatNLO~\cite{Frixione:2002ik} and
\POWHEG~\cite{Nason:2004rx,Frixione:2007vw}, alongside the two
respective computer programs {\tt
  MadGraph5\_aMC@NLO}\xspace~\cite{Alwall:2014hca} and the {\tt
  POWHEG-BOX}~\cite{Alioli:2010xd,Jezo:2015aia}, as well as the
independent implementations within the {\tt
  Herwig}~\cite{Bellm:2015jjp,Reuschle:2016ndi} and {\tt
  SHERPA}~\cite{Hoche:2010pf,Sherpa:2019gpd} event generators.
These and other matching procedures (e.g.~\cite{Bauer:2008qh,Lonnblad:2012ix,Jadach:2015mza}) were
all developed at a time when 
parton showers had only leading logarithmic accuracy, and one of the
questions that was relevant was whether they preserved that leading-logarithmic
(LL) accuracy.

In recent years, a number of developments towards next-to-leading
logarithmic (NLL) accurate parton showers have taken
place~\cite{Dasgupta:2018nvj,Dasgupta:2020fwr,Hamilton:2020rcu,Karlberg:2021kwr,Hamilton:2021dyz,vanBeekveld:2022zhl,vanBeekveld:2022ukn,Forshaw:2020wrq,Holguin:2020joq,Nagy:2020rmk,Nagy:2020dvz,Herren:2022jej},
and one can now investigate the interplay between fixed-order matching
and higher logarithmic accuracy.
To do so, and to help highlight some of the essential considerations
that arise, we focus here on the simplest possible process, NLO
matching for $\gamma^*/Z \to q\bar q$ and $H \to gg$ (in the heavy-top
limit), using the
PanScales family of final-state showers, which all fully satisfy the
broad NLL criteria set out in
Refs.~\cite{Dasgupta:2018nvj,Dasgupta:2020fwr}.

Aside from the obvious requirement that matching should preserve the
(NLL) logarithmic accuracy of a given shower, the main question that
we ask here is whether it can \emph{augment} the logarithmic accuracy,
at least in some situations.
To help understand how this might be the case, it is useful to recall
the traditional resummation formula for two-jet event shapes and jet
rates in two-body decays~\cite{Catani:1992ua}.
Specifically, the probability $\Sigma$ for an observable $O$ to have
a value below some threshold $e^{L}$ (with $L<0$) is given by
\begin{equation}
  \Sigma(O < e^L) = \left( 1+ \frac{\alpha_s}{2\pi} \, C_1 + \ldots\right)
  e^{\as^{-1} g_1(\alpha_sL) + g_2(\alpha_sL) + \as g_3(\as L) + \dots}
   ,\qquad |L| \gg 1\,.
  \label{eq:global-evs-resum-intro}
\end{equation}
The $g_1$ function is responsible for LL terms ($\as^n L^{n+1}$),
$g_2$ for NLL terms ($\as^n L^n$) and both $C_1$ and $g_3$ for NNLL
terms ($\as^n L^{n-1}$).\footnote{To make $C_1$ unambiguous, one can
  define $g_3(0)=0$.}
An alternative way of writing $\Sigma$ is
\begin{equation}
  \Sigma(O < e^L) = h_1(\as L^2) + \sqrt{\as} \,h_2(\as L^2) + \as
  h_3(\as L^2) + \ldots
   ,\qquad |L| \gg 1\,,
  \label{eq:global-evs-resum-intro-DL}
\end{equation}
where the $h_1$ function is responsible for double-logarithmic (DL)
terms ($\as^n L^{2n}$), $h_2$ for NDL terms ($\as^n L^{2n-1}$), $h_3$
for NNDL terms ($\as^n L^{2n-2}$), and so forth.
As is well known in the literature on event shapes, NLL resummation
automatically implies NDL accuracy, and further inclusion of the $C_1$
in Eq.~(\ref{eq:global-evs-resum-intro}) is sufficient to achieve NNDL
accuracy~\cite{Catani:1992ua}.
In analytical resummation, $C_1$ is typically obtained through matching with a NLO
calculation.
The natural question in a parton shower context is therefore whether
matching NLL parton showers with NLO retains the NLL accuracy and
additionally achieves NNDL accuracy for two-jet event shapes and the
two-jet rate.\footnote{In the context of parton-shower merging of
  events with different jet multiplicities, the perspective of
  logarithmic accuracy has already shown its value at NDL
  accuracy~\cite{Catani:2001cc}.}

In examining this question, we will consider three matching
approaches: a straightforward multiplicative matching (similar to the
matrix element corrections of Refs.~\cite{Bengtsson:1986hr,Seymour:1994df},
generalised in Vincia beyond the first order~\cite{Giele:2007di} and
related also to the KrkNLO method~\cite{Jadach:2015mza}), which multiplies the event or splitting weight
by the ratio of the true matrix element to the effective shower matrix
element for the emission of an extra parton;
the \MCatNLO method, which adds in the difference between the true
matrix element and the effective shower matrix element for the
emission of an extra parton;
and the \POWHEG method, which takes full responsibility for the first
emission, and then hands the event over to the shower for the
remaining emissions.
For the first two, retaining NLL and achieving NNDL accuracy for two-jet event shapes
will be relatively straightforward, essentially because the only
kinematic region in which they act is the hard region.\footnote{Since
  the \MAcNLOPS method~\cite{Nason:2021xke} can be viewed as a
  combination of multiplicative and additive (\MCatNLO) matching, we
  also expect it to straightforwardly retain NLL and achieve NNDL accuracy for
  two-jet event shapes, though we defer explicit implementations and
  tests to future work. }
In contrast, because the \POWHEG method always generates the hardest
emission, which is often in the infrared (large-logarithm) region, the
interplay with the shower logarithmic accuracy is more delicate.
Note that understanding the matching/shower interplay of
hardest-emission generator (HEG) methods, like \POWHEG, is important
because such methods also underpin the most widespread NNLO matching
approaches, MiNNLO~\cite{Hamilton:2013fea,Monni:2019whf} and
Geneva~\cite{Alioli:2013hqa} (the other main NNLO matching procedure,
UNNLOPS~\cite{Hoche:2014uhw}, does not fall into this class).

In Ref.~\cite{Frixione:2007vw} it was shown that the
\powheg Sudakov form factor achieves NLL accuracy for the hardest
emission, and in most of this paper we shall take this NLL accuracy of
the \powheg Sudakov as a given.
%
%
It should be clear that if \powheg is then followed by a LL shower,
that NLL accuracy will be squandered for practical
observables.
Less straightforward, however, is the question of what happens when
\powheg, with its NLL Sudakov, is followed by a NLL shower.

The subtleties that we have found in the \POWHEG case are closely associated with
discussions in the literature~\cite{Catani:2001cc,Nason:2004rx,Corke:2010zj}
about
how to connect the effective shower starting scale with the scale of
the \POWHEG emission.
In particular, it is standard to veto the shower branching, so as to
provide the correct relation between \POWHEG and the shower across the
whole of the soft and/or collinear phase space, with no under- or
double-counting.
With the developments of the past few years on parton shower
logarithmic
accuracy~\cite{Dasgupta:2018nvj,Dasgupta:2020fwr,Hamilton:2020rcu,Karlberg:2021kwr,Hamilton:2021dyz,vanBeekveld:2022zhl,vanBeekveld:2022ukn,Forshaw:2020wrq,Holguin:2020joq,Nagy:2020rmk,Nagy:2020dvz,Herren:2022jej},
it is possible to approach these same questions with a range of new
techniques,
both analytical and numerical, that help analyse that
logarithmic accuracy.
In particular, concerns raised at the time of Ref.~\cite{Corke:2010zj}
about shower-induced recoil modifying the first (\POWHEG) emission are
closely connected with the origin of NLL failures discussed in
Ref.~\cite{Dasgupta:2018nvj} and, from a NNDL perspective, can be
solved once one has NLL shower accuracy: the shower can still modify
the first emission, but does so only when the shower emits in the
immediate angular and transverse momentum vicinity of the first
emission.
This ensures that the effect of any modification is NNLL (which also
implies that it is beyond NNDL for event shapes).
In contrast, the questions of avoiding double- and/or under-counting
will relate both with retaining NLL accuracy and the augmentation to
NNDL.

In this paper, we will work with an $e^+e^-$-specific generalisation of
\POWHEG, which employs an ordering variable that coincides in the soft
limit with the generalised PanScales ordering variable,
$v \sim k_t \theta^{\betaps}$, parametrised by $\betaps$.
We will refer to it as \POWHEGbeta.\footnote{Since $\betaps$
  needs to be chosen to be the same in the \POWHEGbeta step and the
  shower, we are explicitly giving up on the property that a single
  run of \POWHEG should be valid for use with any subsequent shower.}
This generalisation enables us to avoid the considerable complications associated with the
need for truncated showers~\cite{Nason:2004rx} and the related subtleties of interplay
with NLL accuracy.
However it leaves open interesting and important questions about the
impact of mismatches in the hard-collinear region.
As we shall see, these mismatches can arise not only from differences in
kinematic maps, but also from the way in which showers 
partition the $g \to gg$ (and potentially $g \to q\bar q$)
splitting functions, which effectively breaks their symmetry.
Both aspects are relevant for NLL/NNDL accuracy.

Note that the ingredients that we discuss here, while sufficient for
event-shape NNDL accuracy, are not enough to obtain NNDL accuracy for
more general observables, e.g.\ for the recently studied sub-jet
multiplicities~\cite{Medves:2022ccw,Medves:2022uii}.

The paper is structured as follows: In section~\ref{sec:matching} we
briefly recall the relevant features of the multiplicative, \MCatNLO
and \POWHEG methods.
In section~\ref{sec:nndl-requirements} we examine how kinematic
mismatches and splitting function de-symmetrisation can affect NNDL
accuracy, with both qualitative explanations and explicit
calculations.
We then verify numerically in section~\ref{sec:log-tests} that
PanScales showers that are suitably matched to NLO do reach NNDL
accuracy for a large set of global event shapes.
In section~\ref{sec:pheno} we briefly investigate the impact of
the matching schemes in a phenomenological context.
We conclude in section~\ref{sec:conclusions}.
Appendix~\ref{sec:appC1} collects the results for $C_1$ coefficients
that we use in Eq.~(\ref{eq:global-evs-resum-intro}),
Appendix~\ref{sec:spin} discusses our treatment of spin correlations
in the hard region, including fixed-order validation plots,
and Appendix~\ref{sec:technical} gives further technical details of
our matching procedures.

\section{Brief overview of standard matching methods}
\label{sec:matching}

For an in-depth review of NLO matching procedures, the reader may wish
to consult Ref.~\cite{Nason:2012pr}, as well as the original papers.
Here we give just a brief overview of the main matching approaches,
with notation adapted from Ref.~\cite{Nason:2021xke} so as to be more
explicit about the starting scale for the showering following the
matched emission.
For simplicity, we will ignore the shower cutoff in all of
our expressions.

\subsection{Multiplicative matching}
We start with the simplest kind of matching, multiplicative
matching.
We write the differential cross section $d\sigma$ in such a way as to make
explicit the structure associated with the first (i.e.\ hardest) emission,
schematically
\begin{equation}
  \label{eq:multiplicative}
  \mathd \sigma_\text{mult} =
  \BB(\PhiB)\,
  \left[
    S_\ps(v_\Phi^\ps,\PhiB)
    \times \frac{\Rps(\PhiBR)}{B_0(\PhiB)}  \, \mathd \PhiBR
    \otimes \frac{R(\PhiBR)}{\Rps(\PhiBR)}\right]
  \times I_\ps(v_\Phi^\ps, \PhiBR)\,.
\end{equation}
Here $\PhiBR$ is the full Born plus one parton phase space, $\PhiB$ is
the underlying Born phase space, and we define the radiation phase
space, $\Phirad$, through $\mathd\Phi = \mathd\PhiB \, \mathd\Phirad$.
In the shower, for a given $\PhiB$ and $\Phirad$, there is an
associated value of a shower ordering variable $v^\ps_{\PhiBR}$.
The Born squared matrix element is given by $B_0(\PhiB)$, the true
Born plus one-parton matrix element is $R(\Phi)$.
The parton-shower approximation to it, $R_\ps(\Phi)$, is expected to
coincide with $R(\Phi)$ in the soft and/or collinear limits.
The parton-shower Sudakov form factor $S_\ps(v,\PhiB)$ is
given by
\begin{equation}
  \label{eq:Sudakov}
  S_\ps(v,\PhiB) = \exp\left[-\int_{v_{\Phi}^\ps>v} \frac{\Rps(\Phi)}{B_0(\PhiB)} \mathd \Phirad\right],
\end{equation}
the NLO normalisation factor can be written\footnote{%
  Given the simplicity of two-body decays (we always take the decaying
  object to be unpolarised), NLO accuracy will not require an explicit
  calculation of the $\BB$ (or, later, $\BB_\ps$) function of this
  section, but can instead be obtained simply by imposing the correct
  overall normalisation of the event sample, e.g.\ for
  $\gamma^*/Z\to q\bar q$ a factor of
  $1 + \frac34 C_F \frac{\as}{\pi}$ relative to the Born cross
  section.  }
\begin{equation}\label{eq:barb}
  \BB(\PhiB)= B_0(\PhiB)+V(\PhiB)+\int R(\Phi) \mathd \Phirad \, ,
\end{equation}
where $V(\PhiB)$ are the virtual corrections and the factor
$I_\ps(v_\Phi^\ps, \PhiBR)$ represents the iterations of the parton
shower branching from a value of the shower ordering variable
$v_\Phi^\ps$ onwards.
Finally, we use the notation $\otimes$ inside the square brackets to
indicate that the first emission is accepted with probability
${R(\PhiBR)}/{\Rps(\PhiBR)}$, and if it is not accepted then the
attempt to create the first emission continues to lower $v_\Phi^\ps$.
This also effectively replaces $\Rps \to R$ in the integrand of the
Sudakov form factor.
The main practical difficulty to be aware of in multiplicative
matching is the requirement ${R(\PhiBR)} \le {\Rps(\PhiBR)}$, in order for
the first emission acceptance probability to be bounded below $1$.%
\footnote{This condition may not always be satisfied, but as long as there are
  no holes in the shower phase space generation, one can generally find
  simple numerical workaround solutions, e.g.\ as discussed in
  Appendix~\ref{sec:technical-panglobal} for the  PanGlobal shower.
 }
From the point of view of logarithmic accuracy, an important feature of
Eq.~(\ref{eq:multiplicative}) is that when $v_\Phi^\ps$ is much
smaller than the shower starting scale, the matching brings no
modifications, because $R_\ps = R$ in that region.

\subsection{\MCatNLO matching}

Next we consider the \MCatNLO procedure, which can be written as
\begin{multline}
  \label{eq:MCatNLO}
  \mathd \sigma_\text{\MCatNLO} =
      \BB_\ps(\PhiB)\, S_\ps(v_\Phi^\ps,\PhiB)
      \times \frac{\Rps(\PhiBR)}{B_0(\PhiB)}  \, \mathd \PhiBR \times I_\ps(v_\Phi^\ps, \PhiBR)
                \,+\\+ \left[R(\PhiBR)-\Rps{}(\PhiBR)\right] \mathd
                \PhiBR \times I_\ps(v^{\max}, \PhiBR) \, ,
\end{multline}
where
\begin{equation}\label{eq:barb-ps}
  \BB_\ps(\PhiB)= B_0(\PhiB)+V(\PhiB)+\int R_\ps(\Phi) \mathd \Phirad \, .
\end{equation}
The interpretation of Eq.~(\ref{eq:MCatNLO}) is that one generates
normal shower events (with a suitable NLO normalisation, $\BB_\ps$) and
supplements them with a set of hard events, with weights distributed
according to $\left[R(\Phi)-\Rps{}(\Phi)\right]$.
This last term vanishes in the infrared (and so has a finite integral)
because $\Rps$ coincides with $R$ in the soft and/or collinear
regions.\footnote{As long as the shower has the correct full-colour
  structure of divergences, which it does in our segment and NODS
  colours schemes~\cite{Hamilton:2020rcu} for the simple cases of
  $\gamma^*$ and Higgs decays.
  Note also that below the shower cutoff, we take
  $\left[R(\Phi)-\Rps{}(\Phi)\right]$ to be zero.
}
This is important for logarithmic accuracy, because as for
multiplicative matching, it ensures that there is no modification to
the showering in the infrared region.
Note that there is freedom in \MCatNLO as to the choice of shower
starting scale in the additional hard events, and here we will use
$v^{\max}$, the largest accessible value. 
Modifying this, e.g.\ taking $v_\Phi^\ps$ instead of $v^{\max}$, only
affects the $\as^2$ contribution in the hard region.
Assuming that this multiplies a double-logarithmic Sudakov associated
with further emissions, it is expected to affect results at
$\as^{n+2}L^{2n}$, which is beyond our target NNDL accuracy.

Our actual implementation of the \MCatNLO procedure will differ
slightly from Eq.~(\ref{eq:MCatNLO}) in terms of how we normalise
different contributions.
The details are given in Appendix~\ref{sec:MCatNLO-appendix}, and the
differences correspond to NNLO corrections that are anyway beyond the
control of NLO matching procedures.

\subsection{\POWHEG matching}
A simple version of the \POWHEG method can be written
\begin{equation}
  \label{eq:powheg-noveto}
  \mathd \sigma_\text{\POWHEG-simple} =
      \BB{}(\PhiB)\, S_\heg(v_\Phi^\heg,\PhiB)
      \times \frac{R_\heg(\PhiBR)}{B_0(\PhiB)}  \, \mathd \PhiBR
      \times I_\ps(v_\Phi^\heg, \PhiBR)\,,
\end{equation}
where the Sudakov form factor $S_\heg$ is defined in analogy with
Eq.~(\ref{eq:Sudakov}) and $R_\heg \equiv R$.
In this simple version of \POWHEG, note the use of $v_\Phi^\heg$ in
$I_\ps(v_\Phi^\heg, \PhiBR)$, i.e.\ the ordering variable is deduced
from the phase space map used with \POWHEG (e.g.\ the
FKS~\cite{Frixione:1995ms} map in the {\tt POWHEG-BOX}), and adopted
directly as a shower starting scale for the remaining shower
emissions.
Throughout this work, when the \POWHEG method is used in conjunction
with a NLL shower, we will assume that the HEG Sudakov factor
$S_\heg(v_\Phi^\heg,\PhiB)$ is also evaluated with NLL accuracy, i.e.\
using two-loop running of $\as$ and the CMW
scheme~\cite{Catani:1990rr}.

In \POWHEG usage with standard transverse-momentum ordered showers,
$v_\Phi^\heg$ in \POWHEG and $v_\Phi^\ps$ in the shower will coincide
for a given phase space point when the emission is simultaneously soft
and collinear.
This is essential in order to achieve leading logarithmic
accuracy.
This is not, however, naturally the case when using $\betaps > 0$ showers in
conjunction with the {\tt POWHEG-BOX}, because of the latter's choice of
transverse momentum as an ordering variable (which corresponds to
$\betaps=0$).
Accordingly, we will adopt a suitably generalised phase space map and
ordering variable that satisfies this property also when used with
$\betaps > 0$ showers, cf.\ Appendix~\ref{sec:appPOWHEG}.
We will call this \powhegbeta.
It will be implicit throughout this
work that the $\beta$ value in \powhegbeta is always taken equal to
$\betaps$.
However, even with \powhegbeta, the two values of $v$ may still differ, for example, when the
emission is collinear and hard.
This creates a mismatch in the infrared between the phase space
covered by the \POWHEGbeta hardest emission generation and the phase space
subsequently covered by the shower.
As we shall see in detail in section~\ref{sec:kinematic-mismatch}, the
mismatch has the potential to create NNDL issues, because the
logarithmic phase space region associated with the mismatch is of
order $1$, the radiation probability in that region comes with one
factor of $\as$ and it can then multiply some part of a double
logarithmic Sudakov factor, giving an overall $\as^{n+1} L^{2n}$.

The aim of Ref.~\cite{Corke:2010zj} was to eliminate the mismatch. 
Given the \POWHEG first-emission event, the recommended approach
within \pythiaeight starts the shower at the maximum allowed scale, but then
for each shower emission $i$, based on the kinematics of $i$, the code
deduces the equivalent value of the ordering variable in the \POWHEG
map, $v_{i}^\heg$, and discards the emission if
$v_{i}^\heg > v_\Phi^\heg$.
We write this procedure as
\begin{equation}
  \label{eq:powheg-veto}
  \mathd \sigma_\text{\POWHEG-veto} =
      \BB{}(\PhiB)\, S_\heg(v_\Phi^\heg,\PhiB)
      \times \frac{R_\heg(\PhiBR)}{B_0(\PhiB)}  \, \mathd \PhiBR
      \times I_\ps(v^{\max}, \PhiBR | v_{i}^\heg < v_\Phi^\heg)\,,      
\end{equation}
where the additional showering condition is indicated after the
vertical bar ($|$) in the $I_\ps$ shower iteration factor.
This should ensure the absence of holes or double-counting in the
infrared phase space (at least when emissions are all well
separated).
With an important proviso concerning the handling of gluon splitting,
discussed below in section~\ref{sec:asym-mismatch}, 
we expect this to be sufficient to restore NNDL accuracy for event
shapes, as long as the underlying shower is NLL accurate.

\section{Matching and event-shape NNDL accuracy}
\label{sec:nndl-requirements}

In this section we will explore how $\order{1}$
phase-space mismatches in the collinear and/or soft region affect NNDL accuracy for event
shapes, in particular in unvetoed \POWHEG style matching,
cf.\ Eq.~(\ref{eq:powheg-noveto}).
We will assume that both the shower and the hardest emission generator
(HEG) have an ordering variable that coincides, in the soft-collinear region, 
with the PanScales family of ordering variables, 
\begin{equation}
v = \frac{k_t}{\rho} e^{-\betaps |\bar{\eta}|} \, ,
\qquad
\rho = \left(\frac{s_{\itilde}
    s_{\jtilde}}{Q^2 s_{\itilde\jtilde}}\right)^{\frac{\betaps}{2}}\,,
\label{eq:panscales-variable-defs}
\end{equation}
parametrised by $\betaps$, and written in terms of the emission
transverse momentum $k_t$ and $\bar \eta$ (which for a back-to-back
dipole is essentially a rapidity), as well as
$s_{\itilde \jtilde} =2\ptilde_{i}\cdot \ptilde_{j}$,
$s_{\itilde}=2\ptilde_{i} \cdot Q$, with $Q$ the total event momentum
and the tildes used to represent pre-branching momenta.
Recall that $\betaps=0$ corresponds to transverse-momentum ordering,
and that PanGlobal showers are NLL accurate for $0\le \betaps < 1$,
while PanLocal showers are NLL accurate for $0< \betaps < 1$.
In Eq.~(\ref{eq:panscales-variable-defs}), $\rho$ is a quantity that is
equal to $1$ for a back-to-back dipole in the event centre-of-mass
frame. 

As well as parameterising the shower evolution variable in terms of
$\betaps$, it is useful to parameterise event-shape observables in
terms of a $\betaobs$ (cf.\ Ref.~\cite{Banfi:2004yd}, where it was
called $b$). Specifically, when the event has just a single soft-collinear emission, the
recursively infrared and collinear safe observable $O$ is given by
\begin{equation}
  \label{eq:O-1-emsn}
  O_{1\text{-emsn}} = \frac{k_t}{Q} e^{-\betaobs |\eta|}\,,
\end{equation}
where $k_t$ is again the transverse momentum of the emission and
$\eta$ is its rapidity with respect to the emitter.
For example, for the thrust~\cite{Brandt:1964sa,Farhi:1977sg}, $\betaobs =1$, and
for the Cambridge $\sqrt{y_{23}}$ 3-jet resolution
scale~\cite{Dokshitzer:1997in}, $\betaobs=0$.
As discussed in the introduction, the quantity we will calculate is
the probability $\Sigma(O < e^{L})$ for
the observable to be below some threshold (where $L$ is negative).

Our discussion will be in two parts.
In section \ref{sec:kinematic-mismatch} we will consider the purely
kinematic differences between the HEG and shower maps, which brings a
logarithmic analysis to the arguments of Ref.~\cite{Corke:2010zj}. 
In section~\ref{sec:asym-mismatch} we will examine additional
important considerations that arise for gluon splitting, connected
with the widespread procedure in parton showers of partitioning the
symmetric $ g\to gg$ and $g\to q\bar q$ splitting functions into two
asymmetric versions, one for each dipole (in the $g \to gg$ case,
each containing just a single soft divergence).

\subsection{Kinematic mismatch between HEG and shower maps}
\label{sec:kinematic-mismatch}

\subsubsection{Qualitative discussion of double-counting}
\begin{figure}
  \centering
  \includegraphics[width=0.8\textwidth]{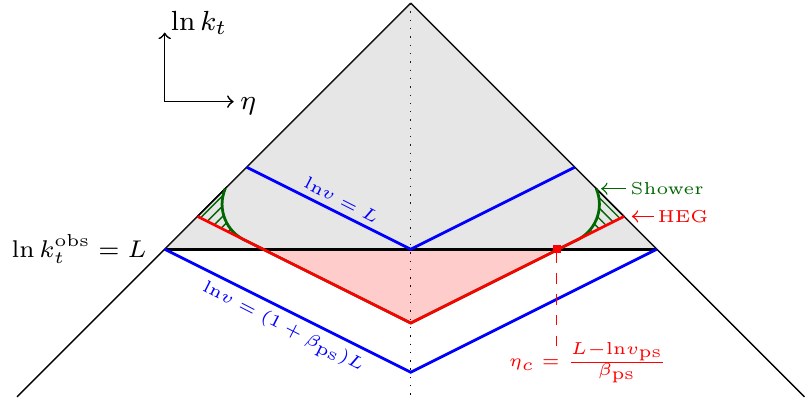}
  \caption{Lund plane~\cite{Andersson:1988gp} representation of the
    phase space for soft and/or collinear emission, and illustration
    of the interplay between a hardest emission generator (HEG), a
    parton shower and a constraint on an observable (black line).
    Dimensionful quantities ($k_t$, $v$) are taken normalised to the
    centre-of-mass energy $Q$.
  }
  \label{fig:nndl-discr-main}
\end{figure}

The case we shall concentrate on is that where $\betaobs=0$ and
$\betaps > 0$.
This is represented in Fig.~\ref{fig:nndl-discr-main}.
The threshold on the observable is represented by the horizontal black
line.
We imagine a shower whose ordering variable coincides with the
HEG in the soft limit, but in the hard-collinear region, for a given
value of the ordering variable, generates a configuration that is
somewhat harder than the HEG (by a factor of order one).
This is illustrated with the green shower contour bending upwards
relative to the red HEG contour.
Though schematic, for simple $e^+e^- \to q\bar q$ events, this picture
captures the essence of a $\betaps=\frac12$ extension of \POWHEG together
with the PanLocal $\betaps=\frac12$ shower.

It is clear that to have $O < e^{L}$, the HEG emission must be below
the upper blue contour in Fig.~\ref{fig:nndl-discr-main}, i.e.\ it
must satisfy $\ln v < L$.
If the HEG emission occurs between the two blue lines (e.g.\ on
the red line), at a rapidity that
places it below the black line (i.e.\ the observable threshold), the
event will contribute to $\Sigma(O < e^L)$ only if the shower does not
emit further above the black line.
That probability is given by the initial HEG Sudakov multiplied by
the part of the remaining shower Sudakov that is above the black line.
The fundamental problem is that the shaded region between the HEG
(red) and shower (green) curves will be accounted for twice, once in
the HEG Sudakov and once in the shower Sudakov, resulting in an
over-suppression of $\Sigma(O < e^{L})$.

More generally we expect the following properties to hold: for
contour mismatches in the hard-collinear region we anticipate NNDL
artefacts for $\betaobs < \betaps $.
We do not expect an NNDL artefact for $\betaobs\ge \betaps$.
In particular, $\betaobs > \betaps $ would correspond to a version of
Fig.~\ref{fig:nndl-discr-main} where the black observable-constraint
contour is steeper than the two blue shower contours, so the mismatch
region is always below the observable-constraint contour and so does
not affect the observable.
An analogous argument can be made concerning potential contour
mismatches in the soft large-angle region, for which we anticipate
NNDL artefacts when $\betaobs > \betaps$.
For the case of $e^+e^- \to \text{2 jets}$, in the context of the
PanScales showers and our specific \POWHEGbeta formulation, there is
no mismatch in this region. 
However, the issue of soft large-angle mismatches becomes important to
address when the Born configuration involves dipoles that are not
back-to-back, for example $e^+e^- \to \text{3 jets}$ or
$pp \to Z+\text{jet}$.
Such configurations are beyond the scope of this work.\footnote{Note
  that in the \pythiaeight shower, the issue arises already with two
  Born partons, because a large-angle soft emission configuration can
  arise from two different values of the ordering variable, cf.\
  Fig.~1 of Ref.~\cite{Dasgupta:2018nvj}.
}

Below, we will give explicit calculations in the context of a double
logarithmic approximation.
Before doing so it is useful, however, to consider why it is that the
HEG/shower combination can achieve NLL accuracy.
Let us assume that the HEG and shower contours align in the soft
and/or collinear region and also that the shower is NLL accurate (as
is the HEG Sudakov).
A crucial point to keep in mind is that the only difference between
running the HEG and then the shower, or instead just the shower, is an
$\order{\as}$ relative correction to the overall normalisation, as
well as the coefficient of the $\order{\as}$ probability of having an
emission in the hard region.
Both are NNLL effects.
Aside from this, the shower continues after the HEG in exactly the
same way as it would had the first emission come from the shower.
It is for this reason that the result maintains NLL accuracy.

One comment is that the agreement of contours can be arranged either
by a vetoing procedure as in Eq.~(\ref{eq:powheg-veto}), or by
adapting the design of the HEG and/or the shower such that their
contours match naturally.
This latter approach, while requiring more tailoring of the HEG/shower
combination, has the advantage that it eliminates complexities
associated with the implementation of the vetoing procedure,
complexities that are likely to add extra challenges as one extends
showers to higher logarithmic accuracy.

\subsubsection{Evaluation of the discrepancy for $\betaobs=0$,
  $\betaps>0$}

To make the argument quantitative, we will essentially work in a
double-logarithmic (DL) approximation for the event shape distribution, and
then specifically evaluate the subleading logarithmic effect of the
mismatch.
We will need the form of the HEG Sudakov, which we write in pure DL
form as
\begin{equation}
  \label{eq:S-heg}
  S_\heg(v) = \exp\left(- \frac{\abar \ln^2 v}{1 + \betaps} \right)\,,
    \qquad
    \abar \equiv \frac{2C\as}{\pi}\,,
\end{equation}
where $C=C_F$ for a $q\bar q$ event and $C_A$ for a $gg$ event, and
the HEG uses the same $\beta$ as the parton-shower $\betaps$. 
As a first step, it is instructive to consider how we obtain the
analogous pure DL form for $\Sigma(O < e^{L})$ in a scenario where there
is no mismatch between HEG and shower contours. 
It is given by 
\begin{subequations}
  \label{eq:pure-HEG}
  \begin{align}
    \label{eq:pure-HEG-line1}
    \Sigma(O < e^{L})
    &=
      e^{-\abar (1+\betaps) L^2}
      +
      e^{-\abar L^2}
      \int_{(1+\betaps)L}^L d\ell
      \frac{2\abar(L-\ell)}{\betaps}
      \exp\left[-\abar
      \frac{(L-\ell)^2}{\betaps} 
      \right],
    \\
    \label{eq:pure-HEG-line2}
    &=
      e^{-\abar L^2}\,,
  \end{align}
\end{subequations}
\logbook{}{where the integrals and sum are done in maths/HEG-shower-mismatch.nb.}%
where the first term in Eq.~(\ref{eq:pure-HEG-line1}) is the
probability for the HEG emission to be below the lower blue contour in
Fig.~\ref{fig:nndl-discr-main},
$\ln v_\heg < (1 + \betaps)L$, i.e.\ it corresponds to
$S_\heg(e^{(1+\betaps)L})$.
The second term is the integral for the HEG emission to be between the
two blue contours, with two additional conditions.
A first condition is that the HEG emission should be on the part of the HEG
(red) contour that is below the observable constraint (black line)
$\ln k_t < L$, which corresponds to a requirement
\begin{equation}
  \label{eq:6}
  |\eta| < |\eta_c| = \frac{L - \ln v_\ps}{\betaps}\,,
\end{equation}
where $\ln v_\ps \equiv \ell$ in Eq.~(\ref{eq:pure-HEG-line1}).
The integral over allowed rapidities gives the first factor in the
integrand of Eq.~(\ref{eq:pure-HEG-line1}).
A second condition associated with this integrand is that subsequent
shower emissions should also be below the observable constraint (i.e.\
there should be no emissions in either the grey-shaded triangle, or the
pink-shaded triangle).
The combination of HEG Sudakov and effective shower Sudakov for this
second condition gives the exponential factors in the second
(integral) term of 
Eq.~(\ref{eq:pure-HEG-line1}).

As a next step, we consider the impact of the mismatch between the
HEG and shower contours, up to NNDL accuracy.
The result is almost identical to Eq.~(\ref{eq:pure-HEG}), 
\begin{subequations}
  \label{eq:shower-HEG}
  \begin{align}
    \label{eq:shower--HEG-line1}
    \Sigma(O < e^{L})
    &=
      e^{-\abar (1+\betaps) L^2}
      +
      e^{-\abar L^2}
      \int_{(1+\betaps)L}^L \!d\ell
      \frac{2\abar(L-\ell)}{\betaps}
      \exp\left[-\abar
      \frac{(L-\ell)^2}{\betaps} 
      -2\abar \Delta
      \right]\!,
    \\
    \label{eq:shower-HEG-line2}
    &=
    e^{-\abar L^2} \left[1 + 2(e^{-\abar \betaps L^2} - 1) \abar
    \Delta + \order{\text{N}^3\text{DL}}\right],
  \end{align}
\end{subequations}
with just an additional term inside the exponent on the first line,
involving a quantity $\Delta$ that represents the effective size of
one mismatch region (shaded green in Fig.~\ref{fig:nndl-discr-main}).
It arises because of the extra shower phase space that needs to be
vetoed when the contours don't match.
For $e^+e^- \to q\bar q$ it reads
\begin{equation}
  \label{eq:Delta-generic}
    \abar \Delta
    =
    \int_0^1 d\zeta \frac{\as C_F}{2\pi} p_{qq}(\zeta)
    \cdot
    2\ln \frac{\theta^\ps_{ik}(v,\zeta)}{\theta^\heg_{ik}(v,\zeta)}\,,
\end{equation}
where $p_{qq}(\zeta) = (1 + \zeta^2)/(1-\zeta)$ is the reduced $q \to
qg$ splitting function and $\theta^\ps_{ik}(v,\zeta)$ is the $\itilde
\to ik$ splitting angle in the parton shower for a given value of the
ordering variable $v$ and post-splitting quark momentum fraction
$\zeta$; $\theta^\heg_{ik}(v,\zeta)$ is the analogue for the HEG.
For the HEG/shower combinations that we consider, the ratio of angles
in Eq.~(\ref{eq:Delta-generic}) will always be independent of $v$. 

A key feature of Eq.~(\ref{eq:shower-HEG-line2}) is that the
correction is indeed NNDL and starts at order $\as^2 L^2$.
Furthermore it vanishes for $\betaps = 0$ (i.e.\ $\betaps = \betaobs$
since we have taken $\betaobs=0$ in our calculation).
This is consistent with our expectation
that the discrepancy is present only for $\betaobs < \betaps$.

To concretely evaluate Eq.~(\ref{eq:Delta-generic}), we need to know
the $\theta_{ik}(v,\zeta)$ functions in the collinear limit.
For the PanLocal and PanGlobal showers~\cite{Dasgupta:2020fwr} and for
our extension of \POWHEG as given in Appendix~\ref{sec:appPOWHEG}, we
have
\logbook{7be1f6909}{Run 2020-eeshower/analyses/test-3jet-matching/test-eqs-3.8}
\begin{subequations}
  \label{eq:thetaik}
  \begin{align}
  \label{eq:thetaik-PL}
    \text{PanLocal:}& \qquad
                      \theta_{ik}(v,\zeta) = 2
                      \cdot \left(\frac{v}{Q}\right)^{\frac{1}{1+\betaps}}
                      (1-\zeta)^{\frac{\betaps}{1+\betaps}} \cdot \frac{1}{\zeta(1-\zeta)}\,,
    \\
  \label{eq:thetaik-PG}
    \text{PanGlobal, \POWHEG{$_\beta$}:}&  \qquad
                                          \theta_{ik}(v,\zeta) = 2
                                          \cdot \left(\frac{v}{Q}\right)^{\frac{1}{1+\betaps}}
                                          (1-\zeta)^{\frac{\betaps}{1+\betaps}} \cdot \frac{1}{1-\zeta}\,.
  \end{align}
\end{subequations}
Note that the equations are identical except in the last factor, which
differs because in PanLocal, the emitter acquires the transverse
recoil, while in PanGlobal that transverse recoil is shared across
through a boost of the whole event (which in the collinear limit
leaves the $ik$ angle unchanged).
In the case where we use PanGlobal or \POWHEG as HEG followed by
the PanLocal shower, combining Eqs.~(\ref{eq:thetaik}) with
Eq.~(\ref{eq:Delta-generic}) results in the following expression
for $\abar \Delta$:
\begin{subequations}
    \label{eq:Delta-PG-PL}
    \begin{align}
    \abar \Delta
    &=
      \int_0^1 d\zeta \frac{\as C_F}{2\pi} p_{qq}(\zeta)
      \cdot
      2\ln \frac{\theta^\ps_{ik}}{\theta^\heg_{ik}}\,,
    \\
    \label{eq:Delta-PG-PL-1}
    &=
      \int_0^1 d\zeta \frac{\as C_F}{2\pi} p_{qq}(\zeta)
      \cdot
      2\ln \frac{1}{\zeta}\,,
    \\
    \label{eq:Delta-PG-PL-2}
    &=
      \frac{2C_F \as}{\pi} \cdot \frac{4\pi^2 -15}{24}\,.
  \end{align}
\end{subequations}
For $H \to gg$, a similar calculation (with values of $w_{gg} = w_{qg} = 0$ for
the parameters governing the de-symmetrisation of the gluon splitting functions,
see section~\ref{sec:asym-mismatch}) gives
\begin{equation}
   \label{eq:Delta-PG-PL-h2gg}
   \abar \Delta =
     \frac{\as}{\pi}\frac{C_A(12 \pi^2 - 49)+8T_R n_f}{36}\,.
\end{equation}

A final comment is that the NNDL-breaking effects discussed here can
also be thought of as a violation of the PanScales conditions that
there should not be long-range correlations between emissions at
disparate rapidities.
Specifically, without vetoing, the presence of a soft-collinear HEG
emission is effectively changing the probability of a hard-collinear
shower emission.
As a result, the HEG/shower combination fails to reproduce the
factorisation that is present in physical matrix elements for
emissions at disparate angles.
This also impacts the analytical resummation structure.
To see how, we take $\Sigma$ from Eq.~(\ref{eq:shower-HEG-line2}), and
extract the highest power $L$ at each order in $\as$, which gives%
\logbook{}{see maths/HEG-shower-mismatch-non-exponentiation.nb, which
  considers result up to $\abar^8$ and from which we deduce the
  general series behaviour.
}
\begin{equation}
  \label{eq:explicit-lnSigma}
  \ln \Sigma = - \abar L^2 - \sum_{n=2}^\infty \frac{2
    \betaps^{n-1}\Delta}{(n-1)!} \cdot \abar^n  L^{2n-2} +
  \order{\abar^n L^{2n-3}}\,.
\end{equation}
The result clearly fails to satisfy the exponentiation property of
Eq.~(\ref{eq:global-evs-resum-intro}), specifically the absence of
terms $\as^n L^m$ in $\ln \Sigma$ with $m > n+1$.
This can be viewed as similar, qualitatively, to the spurious
super-leading logarithms seen in Ref.~\cite{Dasgupta:2020fwr}
(supplemental material, section 3) for standard dipole showers with
observables such as the thrust.
Below, when we summarise matched shower results together with their
logarithmic accuracy, we will use the notation \sout{NLL}, to remind
the reader that the formal NLL accuracy has been lost.
One subtlety, however, is that the difference between
Eqs.~(\ref{eq:shower-HEG-line2}) and (\ref{eq:pure-HEG-line2}) is
always of relative order $\as$.
This has the consequence that in numerical NLL tests with $\as \to 0$
for fixed $\as L$, this difference would mimic a NNLL term, i.e.\ NLL
accuracy would appear to be preserved despite the presence of spurious
super-leading logarithms.

There are, nevertheless, observables that see a larger relative effect.
One example is the invariant mass or transverse momentum of the first
SoftDrop splitting when using
$\beta_\text{SD}=0$~\cite{Dasgupta:2013ihk,Larkoski:2014wba}.
The special characteristic of this observable is that it is not a
standard global event shape, and its resummation does not have
double-logarithmic terms, i.e.\ it starts from $g_2$ in
Eq.~(\ref{eq:global-evs-resum-intro}).
In the fixed-coupling approximation that we have effectively used in
this section, the SD cross section has the following
single-logarithmic structure,
\begin{equation}
  \label{eq:SD-base}
  \Sigma_\text{SD}(L) = e^{\abar c L}\,,
\end{equation}
where $c$ is a constant that depends on SoftDrop's $z_\text{cut}$
parameter, which we take to be small.
Using the same strategy as above, one can explore how
Eq.~(\ref{eq:SD-base}) is modified in HEG/shower combinations with a
hard-collinear mismatch.
Keeping $\betaps = 0$ for simplicity, one finds
\logbook{}{see maths/HEG-shower-mismatch-SD.nb}%
\begin{equation}
  \label{eq:SD-mismatch}
  \Sigma_\text{SD}(L) = e^{\abar c L - \abar \Delta} + e^{-\abar L^2}(1 -
  e^{-\abar \Delta})\,,
\end{equation}
where the coefficient $\Delta$ that parameterises the impact of the
HEG/shower contour mismatch now depends on $z_\text{cut}$.
As with Eq.~(\ref{eq:explicit-lnSigma}), this generates spurious
$\as^n L^{2n-2}$ terms.
If we examine the derivative of $\Sigma_\text{SD}$ (as we will do
below in our phenomenology plots),
\begin{equation}
  \label{eq:SD-mismatch}
  \partial_L \Sigma_\text{SD}(L) =
  \abar c \, e^{\abar c L - \abar \Delta}
  - 2\abar L e^{-\abar L^2}(1 - e^{-\abar \Delta})\,,
\end{equation}
we observe that there is a region, $L \sim 1/\sqrt{\as}$, where the
second term is suppressed relative to the first only by $\sqrt{\as}$.
Thus in this region, the impact of the HEG/shower mismatch is
parametrically larger than the relative $\order{\as}$ correction seen
in Eq.~(\ref{eq:shower-HEG-line2}).

\subsection{Additional subtleties for gluon splitting}
\label{sec:asym-mismatch}

The purpose of this section is to discuss an issue that can arise even
when we have a HEG/shower combination whose kinematic contours (for a
fixed value of the evolution variable) are
aligned not just in the soft-collinear region, but for any
single-emission phase-space point that is soft and/or collinear.
The issue is connected with the fact that the $g \to gg$ splitting
function
\begin{equation}
  \label{eq:pgg-complete}
  \frac1{2!} P_{gg}(\zeta) = C_A \left(\frac{\zeta}{1-\zeta} +
    \frac{1-\zeta}{\zeta} + \zeta(1-\zeta)\right)\,,
\end{equation}
has two soft divergences, one for $\zeta \to 0$ and the other for
$\zeta \to 1$.
This is a consequence of the inherent symmetry
$\zeta \leftrightarrow (1-\zeta)$, which stems from the fact that 
$g \to gg$ corresponds to splitting to two identical particles (hence
also the $1/2!$ factor).
However, dipole showers break this symmetry, through the concept of an
emitting particle (the ``emitter'') and a radiated particle.
In particular, in order to help reproduce the correct pattern of
large-angle soft radiation, dipole showers de-symmetrise the splitting
function so that there is a divergence only when the radiated gluon
becomes soft.
For example the PanScales showers use
\begin{equation}
  \label{eq:pgg-asym}
  \frac1{2!} P_{gg}^{\text{asym}}(\zeta) = C_A \left[
    \frac{1 + \zeta^3}{1-\zeta} + (2\zeta - 1)w_{gg}
  \right]\,,
\end{equation}
where the choice of the $w_{gg}$ parameter fixes arbitrariness
that arises in partitioning the finite part of the splitting
function.
It is straightforward to verify that $P_{gg}^{\text{asym}}(\zeta) +
P_{gg}^{\text{asym}}(1-\zeta) = 2P_{gg}(\zeta)$.

\begin{figure}
  \centering
  \begin{subfigure}{.5\textwidth}
    \centering
    \includegraphics[width=1.\linewidth]{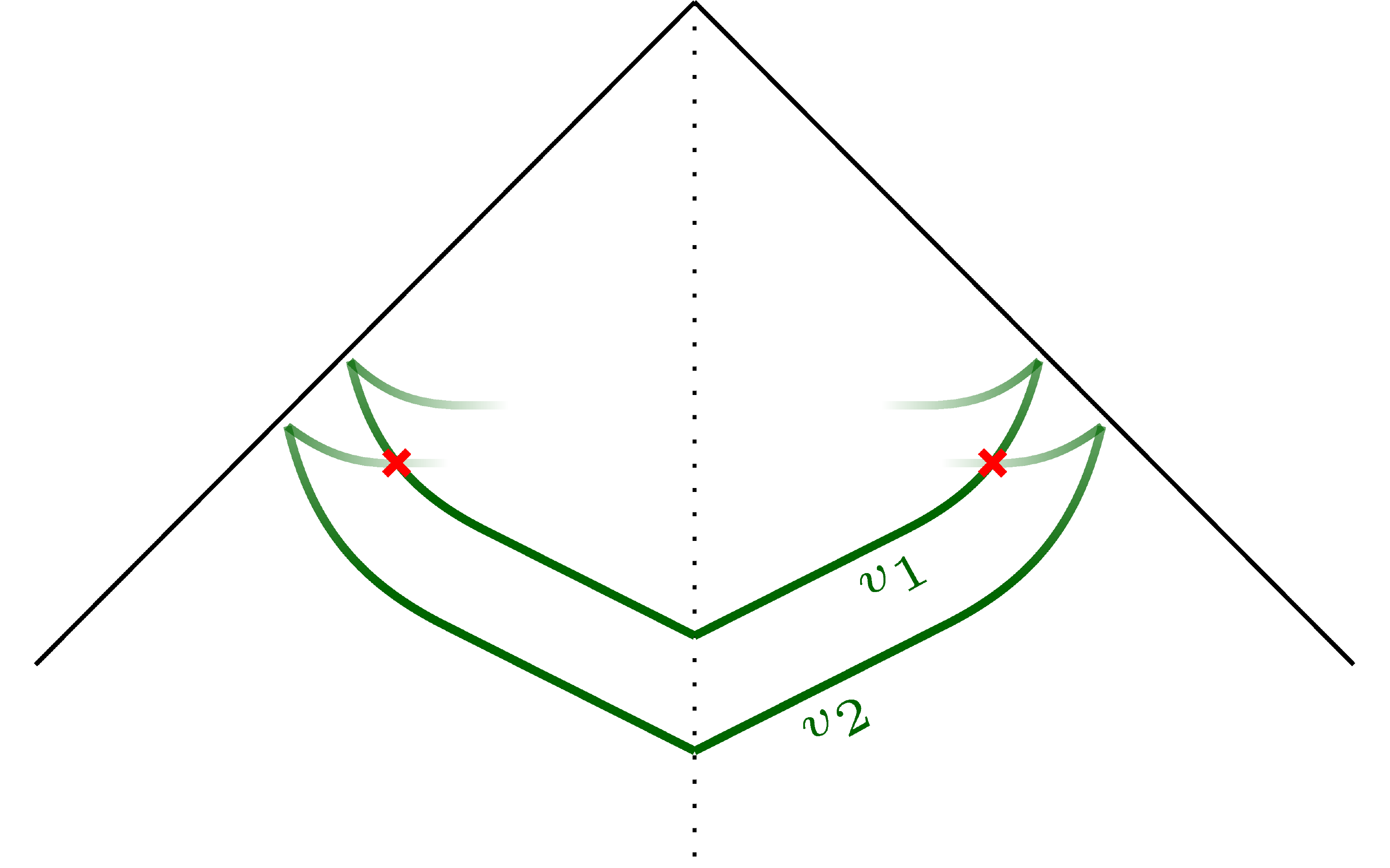}
    \caption{}
  \end{subfigure}%
  \begin{subfigure}{.5\textwidth}
    \centering
    \includegraphics[width=1.\linewidth]{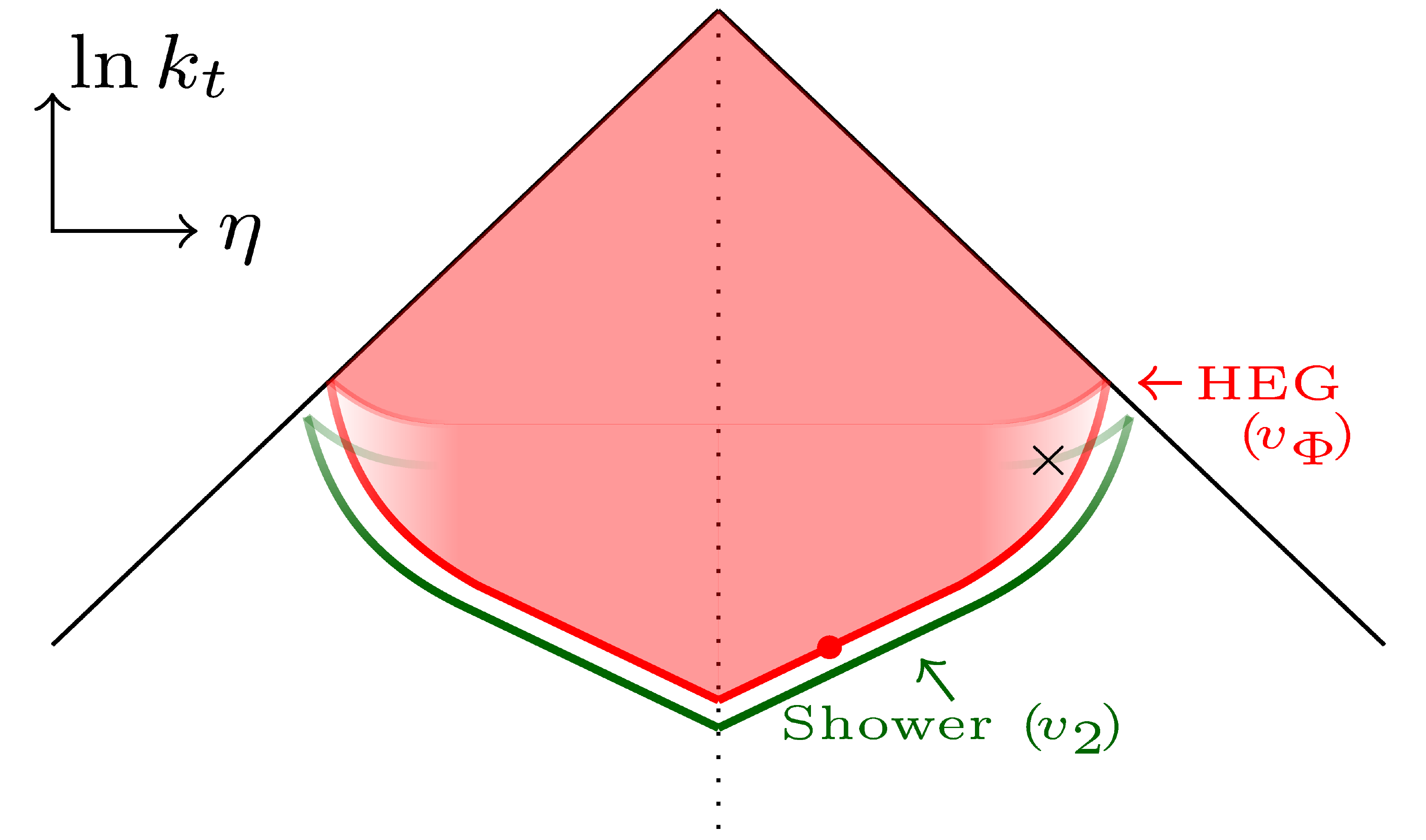}
    \caption{}
  \end{subfigure}

  \caption{Schematic illustration of the issue associated with gluon
    asymmetrisation.
    (a) Contours on the Lund plane, in the PanLocal
    family of showers, highlighting the fact that a given physical
    point $X$ in the Lund plane (highlighted with a red cross) can come from
    two different values of $v$. The shading of the green curves represents
    the variation in radiation intensity along the contour.
    (b) Density plot, at each point in the Lund plane, representing
    schematically the fraction of 
    the emission intensity at that point that has been excluded once
    the HEG has reached a given $v$ value ($v_\Phi$) without
    emitting, and an illustration that as the shower continues there
    may still be phase-space points (such as that marked with a cross) where the
    Sudakov has only been partially accounted for.
    The implications are discussed in the text.
    \label{fig:gluon-asymmetrisation}
  }
\end{figure}

The hard matrix element generated by the HEG can be de-symmetrised
similarly.
The {\tt POWHEG-BOX} code follows the FKS procedure~\cite{Frixione:1995ms},
which introduces so-called $\mathcal{S}$-functions that are used to partition
the soft and collinear singularities.
The de-symmetrisation discussed above is handled by
an additional multiplicative factor $h(\zeta)$, cf.~Eqs.~(2.76)--(2.77) of
Ref.~\cite{Frixione:2007vw}, with $\zeta$ for an $\itilde \to ik$
splitting defined as $E_i/(E_i + E_k)$.
One can implement the scheme of Eq.~\eqref{eq:pgg-asym} by setting
\begin{equation}
  h(\zeta) = \frac{P_{gg}^{\text{asym}}(\zeta)}{P_{gg}(\zeta)}.
\end{equation}
The reason that the de-symmetrisation matters is that in many cases
the kinematic map is not symmetric under
$\zeta \leftrightarrow (1-\zeta)$.
This can be seen in Eqs.~(\ref{eq:thetaik}), where the only
combination that is symmetric is the PanLocal map for $\betaps = 0$
(this, however, is not NLL accurate).
The asymmetry is illustrated for PanLocal $\betaps=\frac12$ in
Fig.~\ref{fig:gluon-asymmetrisation}a, which shows fixed-$v$ contours 
in the physical Lund plane.
Specifically for emission of $k$ from an $\itilde\jtilde$ Born event,
i.e.\ $\itilde \jtilde \to ijk$, the plot represents
$\ln k_t = \ln [\min(E_i, E_k)\sin \theta_{ik}]$ versus
$\eta = -\ln \tan \theta_{ik}/2$ in the right-hand Lund plane, and
analogously in the left-hand plane with respect to $j$. 
A given contour has two parts: the lower branch corresponds
to $\zeta>\frac12$, and contains the soft divergence; the
upper branch corresponds to $\zeta < \frac12$ and is free of any
soft divergence, so it contributes significantly only in the hard
collinear region.

The critical observation is that any specific point $X$ in the
hard-collinear region of the Lund plane can be populated by two
distinct values of $v$: first at some $v_1$ by the lower
($\zeta>\frac12$) part of the $v_1$ contour, and then at a smaller value
of $v=v_2 < v_1$ by the upper ($\zeta<\frac12$) part of the $v_2$
contour. 
The relative fraction of radiation intensity from the two values of
$v$ at a given point in the Lund plane depends on how the splitting
function has been de-symmetrised, for example on the value of $w_{gg}$
in Eq.~(\ref{eq:pgg-asym}).
For a given shower or HEG, we will refer to $f_{X,1}(w_{gg})$ as the
fraction coming from the $v_1$ contour and
$f_{X,2}(w_{gg}) = 1 - f_{X,1}(w_{gg})$ as the fraction from the $v_2$
contour.

Fig.~\ref{fig:gluon-asymmetrisation}b illustrates how this is relevant
in the combination of HEG and shower.
Suppose that we have a HEG with some $w_{gg} = w^\heg$ and a shower with
some $w_{gg} = w^\ps$.
We consider a situation where the HEG generated the first emission at
some $v_\Phi$, and where that emission is deep into the soft-collinear
region, so that the branching should have no direct kinematic impact on
subsequent hard collinear radiation.
Next, let us focus on the point labelled with a cross (``$X$'').
That point is associated with two values of the evolution variable,
$v_1$  and $v_2$ (remember, the map from evolution variable to
kinematic contour is identical between HEG and shower across the full
soft and/or collinear phase space).
The first value, $v_1$ is already excluded by virtue of the fact that the HEG's
emission corresponded to a smaller value of $v$, i.e.\ $v_\Phi <
v_1$.
In effect some fraction $f_{X,1}(w_\heg)$ of the total Sudakov for not
emitting at $X$ was the HEG's responsibility, and that fraction
depends on the de-symmetrisation parameter $w^\heg$ of the HEG.
The second value of the evolution variable, $v_2$, that is associated
with kinematic point $X$ is the shower's responsibility, since
$v_2 < v_\Phi$.
If the shower does not emit there, the fraction of the total Sudakov
that the shower contributes is $f_{X,2}(w^\ps) \equiv 1 -
f_{X,1}(w^\ps)$.
If $w^\heg = w^\ps$ then those two fractions add up to one.
Otherwise they may be larger or smaller than one, which is equivalent
to having a partial double- or under-counting of the radiation
intensity at $X$, or more generically in the hard-collinear vicinity
of the $v_\Phi$ contour.

The impact of the double counting will be similar to that of the
kinematic-contour mismatch discussed in
section~\ref{sec:kinematic-mismatch}.
Indeed all that needs doing to evaluate its NNDL effect is to take
Eq.~(\ref{eq:shower-HEG-line2}) and replace 
the expression for $\abar \Delta$ in Eq.~(\ref{eq:Delta-generic}) with
\begin{equation}
  \label{eq:Delta-generic-gluon-asym}
  \abar \Delta = \int_0^{\frac12} d\zeta
  \left[ \frac1{2!} P_{gg}^{\ps}(\zeta) - \frac1{2!} P_{gg}^{\heg}(\zeta) \right]
  \cdot
  2\ln \frac{\theta_{ik}(v,\zeta)}{\theta_{ik}(v,1-\zeta)}\,,
\end{equation}
\logbook{}{see maths/HEG-asym-mismatch-H2gg.nb}%
where $\theta_{ik}(v,\zeta)$ is the same function for both the HEG and
the parton shower.
To understand Eq.~(\ref{eq:Delta-generic-gluon-asym}), keep in mind
that $0<\zeta<\frac12$ is the region corresponding to the upper
part of the contour in Fig.~\ref{fig:gluon-asymmetrisation}.
If $P_{gg}^{\ps}(\zeta) > P_{gg}^{\heg}(\zeta)$ then the combination
of HEG and shower is enhancing the Sudakov in that region, much like
the double counting of Fig.~\ref{fig:nndl-discr-main}.
The impact of the double counting on the event shape is structurally similar
to that of the mismatch of contours in section~\ref{sec:kinematic-mismatch}.
It can in fact be thought of as a weighted mismatch of contour, moving
the mismatched part of the weight from $\theta_{ik}(v,1-\zeta)$ to
$\theta_{ik}(v,\zeta)$, from which one deduces
Eq.~(\ref{eq:Delta-generic-gluon-asym}). 

Note that there is a similar consideration for $g \to q\bar q$
branchings, where the splitting function is sometimes also
de-symmetrised. 
For completeness, using the PanScales form for the de-symmetrised
$g \to q\bar q$ splitting,
\begin{equation}
  \label{eq:pqg-asym}
  P_{qg}^{\text{asym}}(\zeta) = T_R n_f \left[2 \zeta^2 - (2\zeta - 1)w_{qg}
  \right]\,,
\end{equation}
we have
\begin{subequations}
  \label{eq:asym-mismatch-panscales}
  \begin{align}
    \label{eq:asym-mismatch-panglobal}
    \abar \Delta^\text{PanGlobal}
       &=
         \left[
            \left(w^{\heg}_{gg} - w^\ps_{gg}\right)C_A
          - \left(w^{\heg}_{qg} - w^\ps_{qg}\right) n_f T_R
         \right]
         \frac{-1}{1 + \betaps}\,,
    \\
    \label{eq:asym-mismatch-panlocal}
    \abar \Delta^\text{PanLocal}
       &=
         \left[
            \left(w^{\heg}_{gg} - w^\ps_{gg}\right)C_A
          - \left(w^{\heg}_{qg} - w^\ps_{qg}\right) n_f T_R
         \right]
         \frac{\betaps}{1 + \betaps}\,,
  \end{align}
\end{subequations}
Note that we will usually choose $w^{\heg}_{gg}=w^{\heg}_{qg}\equiv w^{\heg}$ and
$w^{\ps}_{gg}=w^{\ps}_{qg}\equiv w^{\ps}$, which leads to an almost complete
cancellation between the $C_A$ and $n_f T_R$ terms for $n_f=5$ (the
cancellation would be exact for $n_f=6$).
Accordingly, when we come to test the above analysis numerically with the full
shower below, we will use $n_f=0$ so as to
avoid this cancellation.

As with the kinematic mismatch of
section~\ref{sec:kinematic-mismatch}, the effect that we have just
seen corresponds to a violation of the PanScales conditions that
there should not be long-range correlations between emissions at
disparate rapidities, i.e.\ the presence of a soft-collinear emission
(from the HEG) modifies the probability for subsequent hard-collinear
emission (from the shower).
This, again, has implications for exponentiation.

\subsection{Practical implementation of vetoing}
\label{sec:vetoing-alg}
For completeness, we give here the specific algorithm that we adopt to
achieve the vetoing of parton shower emissions following on from a
first HEG step.
The combination in which we will use it is with a \powheg or
PanGlobal-like HEG, followed by a PanLocal shower (dipole or
antenna).
This combination has the property (cf.\ Eq.~(\ref{eq:thetaik})) that
we only have to deal with double-counting, never with holes in soft
and/or collinear phase space.
The algorithm comes in two parts.

One part keeps track of the indices of the particles that should be
considered the descendants of the particles in the Born configuration
$\PhiB$.
For our cases later of $e^+e^- \to q\bar q$ and $H \to gg$, we label
the Born partons $a$ and $b$.
For any branching (HEG or shower) where the emitter is one of
the partons currently labelled as Born, e.g.\ $\itilde \to ik$ with
$\itilde$ a Born particle, the more energetic of $i$ and $k$
inherits the Born label.
Note that for a $q \to qg$ splitting we could alternatively have
considered the descendent quark to always be the Born particle.
We will discuss the relevance of the choice below.  

The second part of the algorithm follows the various HEG and shower
steps:
\begin{enumerate}
\item Allow the HEG to generate the first emission, resulting in a
  Born$+1$ configuration $\Phi$, with a value $v_\Phi^\heg$ of the HEG
  ordering variable.
  The HEG step updates the Born indices as per above.
\item Start the parton shower with $v^\ps = v_\Phi^\heg$. This is
  allowed because our HEG/shower combinations can lead to double
  counting, but not holes in the soft and/or collinear phase space.
\item For all subsequent emissions $k$ from a dipole
  $\itilde \jtilde $, if the emitter $\itilde$ carries a Born label,
  check the following veto condition:\label{enum:vetostep}
  \begin{enumerate}
  \item Compute $k_t = E_k \sin \theta_{ik}$ and $\eta = -\ln \tan
    \theta_{ik}/2$ (this is done using exact momenta).
  \item Compare these coordinates with the corresponding contour of
    the HEG at $\ln v_\Phi^\heg$, i.e.\
    $\ln k_t^\heg(v_\Phi^\heg,\eta) = \ln v_\Phi^\heg + \betaps|\eta|$ for
    {\powheg}$_\beta$ and for PanGlobal as a HEG.
    \label{enum:vetostep-contour}
  \item If the emission is above the contour,
    $\ln k_t > \ln k_t^\heg(v_\Phi^\heg,\eta)$, veto the emission.
  \end{enumerate}
  For antenna showers, where the emitter/spectator distinction is
  absent, the same check is performed for either end of the dipole,
  insofar as it has the Born label.
\item If the splitting is accepted, update the Born labels.
\end{enumerate}
A few comments may be helpful.
The first concerns the freedom in how we assign the Born label.
For NNDL accuracy, the critical element is that when Born-labelled
parton $\itilde$ splits as $\itilde \to ik$, then when $k$ is a soft
gluon, it should be $i$ that acquires the Born label.
When $k$ is a hard gluon, subsequent vetoing only affects triple-collinear
configurations (and their associated virtual corrections), i.e.\
configurations with three partons at commensurate angles and with
commensurate energies.
Those do not play a role at NNDL accuracy.
A second comment concerns the shower starting scale, which we could equally well
have taken to be $v^\ps = v^{\max}$, as
written in Eq.~(\ref{eq:powheg-veto}), as long as emissions from
non-Born partons are only allowed for $v^\ps < v_\Phi^\heg$.
This should not affect NNDL accuracy, and we have verified that in a
phenomenological context the impact is small, at the percent level.
\logbook{272ac426d7d24}{Plots with a starting $v^\ps = Q$ (``power shower'') are to be
    found in
    2020-eeshower/analyses/global-obs-pheno/matching-paper-runs-LS-17012023-mZ/3jet-column-summary.withpowershower.pdf
    and
    2020-eeshower/analyses/global-obs-pheno/matching-paper-runs-LS-17012023-mZ/3jet-column-summary.wopowershower.pdf
}%
Were we to consider HEG/shower combinations that result in IR holes,
there would be less freedom in the choice of shower starting scale and
it might well be necessary to start with $v^\ps = v^{\max}$.
Finally, it is useful to be aware that the kinematic variables that we
adopt for the contour check in step~\ref{enum:vetostep-contour}
differ from the Lund contours represented in
Fig.~\ref{fig:gluon-asymmetrisation}, specifically with the use of
$E_k$ rather than $\min(E_i, E_k)$.
Insofar as the meaning of $\zeta > \frac12$ is the same in the HEG and
the shower, this helps avoid complications with the folding of
contours seen with the $\min(E_i, E_k)$ choice in
Fig.~\ref{fig:gluon-asymmetrisation}.
However it is still necessary, in the $g \to gg/ q\bar q$ case, to
ensure that the same de-symmetrisation of the splitting functions is
used in the HEG and shower steps.

\section{Logarithmic tests}
\label{sec:log-tests}

\begin{table}
  \centering
  \begin{tabular}{lcccc}
    \toprule
     & mult. & \MCatNLO & \POWHEG{$_\beta$} & HEG$_\text{PanGlobal}$
    \\\midrule
    PanLocal $\betaps=0.5$ (dip.) &\checkmark &\checkmark&\checkmark$_{\!\!(v)}$&
    \\
    PanLocal $\betaps=0.5$ (ant.) &           &\checkmark&          &\checkmark$_{\!\!(v)}$
    \\
    PanGlobal $\betaps=0.0$       &\checkmark &\checkmark&\checkmark&
    \\
    PanGlobal $\betaps=0.5$       &\checkmark &\checkmark&\checkmark&
    \\
                                       \bottomrule
  \end{tabular}
  \caption{The main matching and shower combinations for which we will
  test NNDL accuracy, for both $\gamma^*/Z \to q\bar q$ and $H \to
  gg$.
  A $(v)$ next to a check mark indicates that we use the vetoing
  algorithm of section~\ref{sec:vetoing-alg}.
  We use the NODS colour scheme from Ref.~\cite{Hamilton:2020rcu}.
}
  \label{tab:matching-combos}
\end{table}

We now turn to logarithmic tests with event shapes.
There are quite a few potential combinations of matching scheme and
shower.
The subset that we explore is listed in
Table~\ref{tab:matching-combos}.
The combinations that do not have a check mark would also have been
straightforward to test,\footnote{With the exception of the
  multiplicative scheme for PanLocal (antenna), whose implementation
  is somewhat more complex, because of the existence of physical
  kinematic points that can be reached through three possible
  branching histories.}  and are left out simply to limit CPU usage
and bookkeeping.

We have run the standard NLL tests as in
Refs.~\cite{Dasgupta:2020fwr,Hamilton:2020rcu} at $< 1\%$ target
accuracy, to verify that the matched showers continue to reproduce
full-colour NLL accuracy for global event shapes (and NDL for
multiplicity at $\lesssim 2\%$ target accuracy).
These tests were successful.
Given the large number of results, we refrain from showing them.
Instead, here we concentrate on the NNDL accuracy tests.

For the NNDL predictions we take the standard NLL formulae as used in
earlier work and supplement them with the $C_1$ coefficients as given
in Appendix~\ref{sec:appC1}, some taken from the literature, some
extracted numerically for this work. 
Slightly adapting the notation of the introduction, we denote by
$\Sigma(\as, L)$ the probability for an event shape observable to have
a value $O$ that is less than $e^{L}$ for a given $\as(Q)=\as$
where $Q$ is the $\gamma^* \to q\bar q$ centre-of-mass energy, or
the Higgs-boson mass for $H \to gg$ tests.
For a matched shower to qualify as being accurate at the NNDL level, its
prediction for the cumulative distribution of a given observable,
$\Sigma_\text{PS}$, must clearly satisfy the following criterion,
\begin{equation}
  \lim_{\substack{\alpha_s \to 0 \\ \xi\, \text{fixed}}} 
  \frac{\Sigma_\ps(\alpha_s, -\sqrt{\xi/\alpha_s})
    -
    \Sigma_{\rm NNDL}(\alpha_s,-\sqrt{\xi/\alpha_s})}%
  {\alpha_s \Sigma_{\rm DL}} = 0\,,
\label{eq:NNDL_tests}
\end{equation}
where fixing $\xi$ is equivalent to fixing $\as L^2$, as is relevant
for isolating different terms in the DL-type expansion of
Eq.~(\ref{eq:global-evs-resum-intro-DL}).

To perform the tests, we run the showers with up to six values of
$\alpha_s = \frac{0.1}{N^2}$, with
$N \in \lbrace 3,4,5,6,8,12 \rbrace$.
We will show results at $\xi=\alpha_s L^2 = 1.296$ for
$\gamma^* \to q \bar q$, and $\xi=\alpha_s L^2 = 0.791$ for
$H \to gg$, which correspond to values of the cumulative distributions
$0.25 \leq \Sigma \leq 0.6$, i.e.\ in the bulk of the distribution for
all observables under scrutiny.
In each run, we choose a shower cutoff such that showering continues
substantially below the smallest value needed to accurately predict
the observable at the given $\xi$ value. 
The limit in Eq.~(\ref{eq:NNDL_tests}) is extracted numerically by
extrapolating a fit that is linear or quadratic in powers of
$\sqrt{\as}$.
In $\gamma^* \to q\bar q$ events, we use a linear polynomial fit to
the three smallest $\as$ values. We repeat the fit with the four
smallest $\as$ values and take the difference in the intercept of both
fits as a systematic uncertainty for the extrapolation procedure.
In $H\to gg$ events, we find that there is still a visible quadratic
component for some observables at the $\as$ and $\xi$ values we are considering,
and thus fit a quadratic polynomial to all $\as$ values
(all but the largest for the systematic uncertainty).

We will consider a range of observables with different values
of $\beta_\text{obs}$, cf.\ Eq.~(\ref{eq:O-1-emsn}).
Given the discussion of section~\ref{sec:nndl-requirements}, which
showed that the presence of potential NNDL artefacts depends in a
non-trivial way on the choice of both $\betaobs$ and $\betaps$, it is
important to ensure that for each type of matching, we have
shower/event-shape combinations with $\betaobs < \betaps$,
$\betaobs = \betaps$ and $\betaobs> \betaps$.
The observables that we take are
\begin{itemize}
\item The Cambridge $\sqrt{y_{23}}$ resolution
  parameter~\cite{Dokshitzer:1997in}, total and wide jet broadenings
  $B_T$ and $B_W$~\cite{Catani:1992ua}, which all have $\betaobs=0$
  (i.e.\ the same LL and NDL structures), but differ at NLL and NNDL.
\item The thrust~\cite{Brandt:1964sa,Farhi:1977sg}, and the
  $C$-parameter~\cite{Ellis:1980wv,Parisi:1978eg,Donoghue:1979vi}.
  These both have $\betaobs = 1$, and are equivalent up to NLL
  (considering $C/6$ versus $1-T$), but not at NNDL.
\item Three sets of parameterised observables for
  $\betaobs = 0, \frac12,1$: the fractional moments
  ${\rm FC}_{1-\beta_\text{obs}}$ of the energy-energy
  correlations~\cite{Banfi:2004yd}, as well as the sum and maximum,
  $\sum_i k_{T,i}/Q e^{-\beta_\text{obs} |\eta_i|}$ and
  $\max_i k_{T,i}/Q e^{-\beta_\text{obs} |\eta_i|}$ respectively,
  among primary Lund declusterings
  $i$~\cite{Dreyer:2018nbf,Dasgupta:2018nvj}.
  The first two observables are equivalent to each other at NLL accuracy but
  differ at NNDL, while the latter differs also in its NLL terms.
\end{itemize}

\begin{figure}[h]
  \centering
  \includegraphics[width=.925\textwidth,page=6]{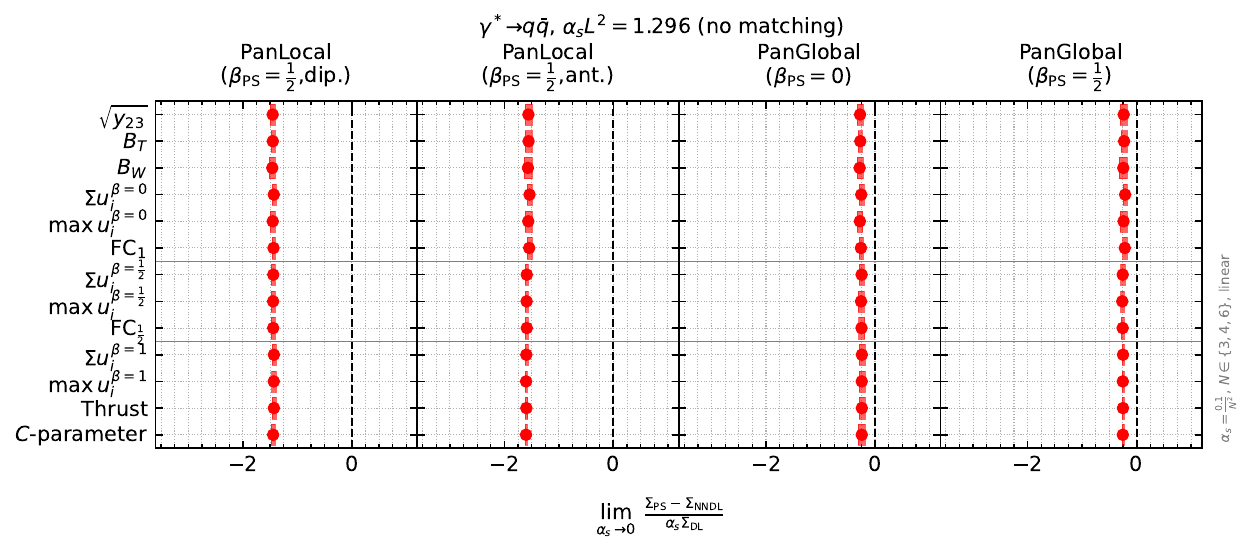} \\
  \includegraphics[width=.925\textwidth,page=18]{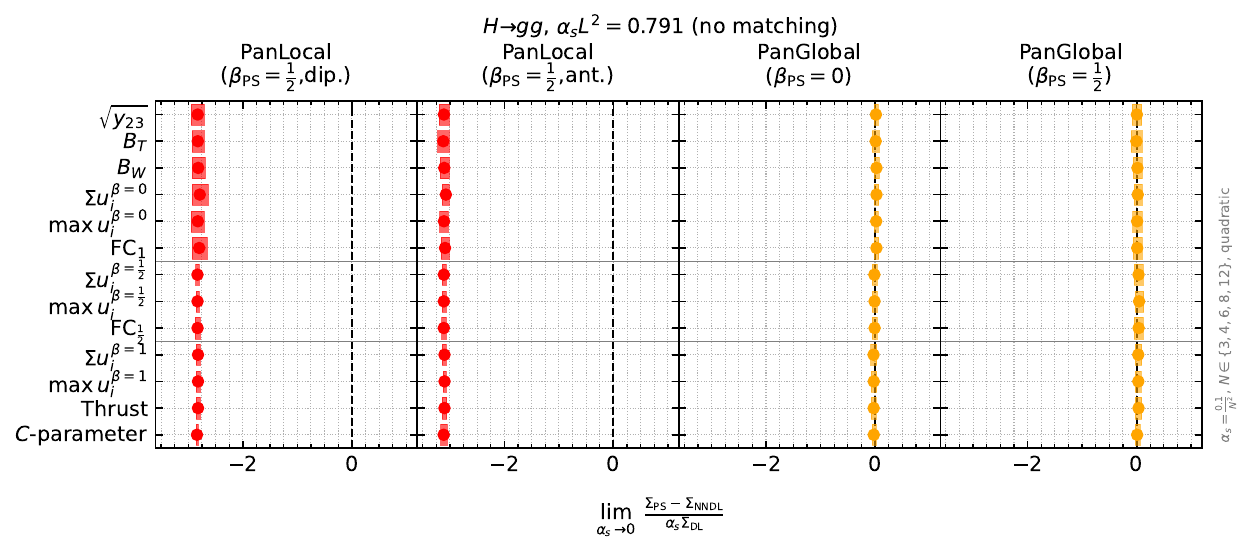}
\caption{Results of the NNDL accuracy tests at fixed $\xi = \as L^2$ for the PanLocal dipole and
  antenna (with $\betaps = \frac12$) and PanGlobal ($\betaps = 0$ and $\betaps = \frac12$)
  showers, without 3-jet matching, for $\gamma^*\to q\bar q$ (top) and $H \to gg$
  (bottom).
  In these and subsequent plots, points marked green (red) show agreement
  (disagreement) with the NNDL results at the $2\sigma$-level. Amber
  points manifest coincidental agreement for $n_f=5$ as explained in the text.
  The plots of this figure have no green points.
}
  \label{fig:NNDL-no-matching}
\end{figure}

We start by showing results for Eq.~(\ref{eq:NNDL_tests}) for showers
without matching, in order to gauge the size of the NNDL discrepancy.
This is shown in Fig.~\ref{fig:NNDL-no-matching}. 
Points are coloured in green if the central value is consistent with
zero within $2\sigma$ (and in red otherwise),
where the $1\sigma$-uncertainty band is given by the
statistical uncertainty and a systematic fit uncertainty added
linearly.\footnote{Note that in contrast to earlier PanScales
  work where statistical and systematic uncertainties were added
  in quadrature, this is a looser criterion. This choice reflects the
  significantly larger number of tests being performed here, and the
  correspondingly higher chance that at least one ``correct''
  shower--matching scheme combination is mislabelled as having
  failed.
  Additionally, as in earlier work, in situations where the tests
  yield a result that is expected to be consistent with a given
  logarithmic accuracy but differs by more than $2\sigma$, we extend
  the runs (either with further statistics or additional $\as$ values)
  so as to clarify whether there is a genuine failure or not.
}
With the exception of the PanGlobal shower in $H \to gg$, all showers
without matching are clearly inconsistent with the NNDL result and
the discrepancy can be significant, notably for the PanLocal showers
where it is of order $2{-}$3.
The PanGlobal $H \to gg$ results are marked in amber, because the
agreement is fortuitous: while the shower's effective 3-jet
matrix-element is different from the exact result, we found that a
coincidental cancellation leads to a seemingly correct result at
$n_f = 5$.\footnote{
  We have also performed runs with $n_f = 0, 9$ and
  verified that there is a non-zero NNDL discrepancy, which coincides
  with expectations from a simple semi-analytic calculation.
  \logbook{e7b1a5879}{See logbook/2022-09-22-shower-C1-coefficients
  and analyses/event-shapes-nndl-tests/results/H2gg_nf-vars}
}
For each shower, the discrepancy is independent of the observable,
because the test is carried out for small values of the observable
(since $\as L^2$ is fixed with $\as \to 0$), while the mismatch
between the effective shower matrix element and the true matrix
element is limited to the hard region.

\begin{figure}[h]
  \centering
  \includegraphics[width=.755\textwidth,page=8]{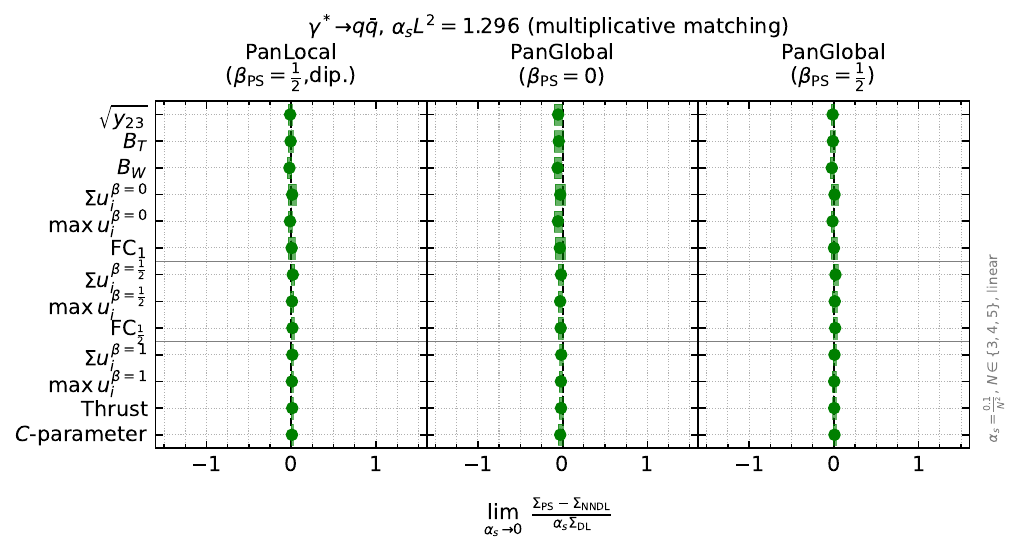}\\
  \includegraphics[width=.755\textwidth,page=20]{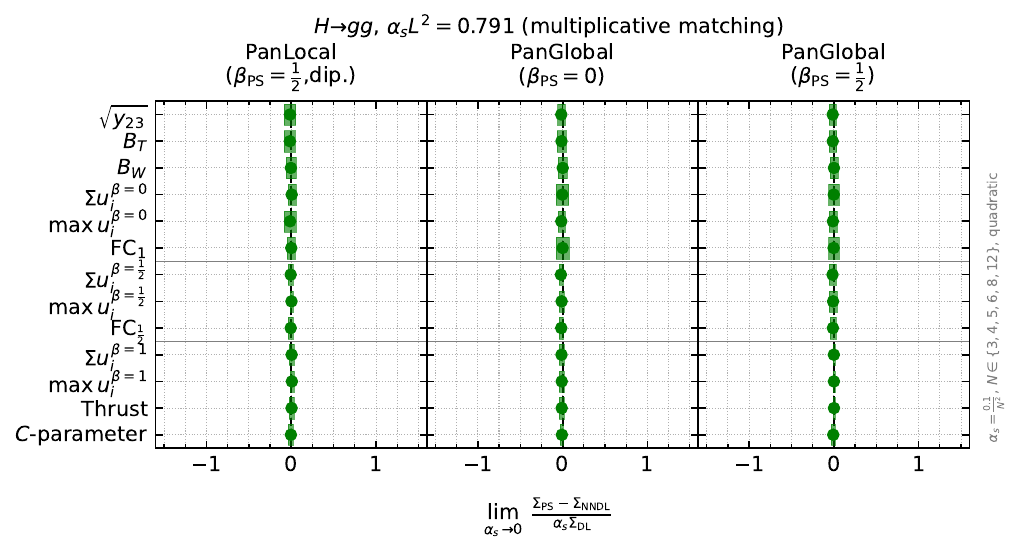}
  \caption{Results of NNDL accuracy tests for the PanLocal dipole ($\betaps=\frac12$)
  and PanGlobal ($\betaps =0,\frac12$) showers, matched with the multiplicative scheme.}
  \label{fig:NNDL-powheg-1}
\end{figure}

Next, we examine results of the matching with the multiplicative scheme.
Fig.~\ref{fig:NNDL-powheg-1} displays the results of NNDL tests for the PanLocal
dipole ($\betaps = \frac12$) and the PanGlobal ($\betaps = 0, \frac12$) showers matched
multiplicatively. These are all in agreement with the NNDL result for all
observables.

\begin{figure}[h]
  \centering
  \includegraphics[width=.925\textwidth,page=10]{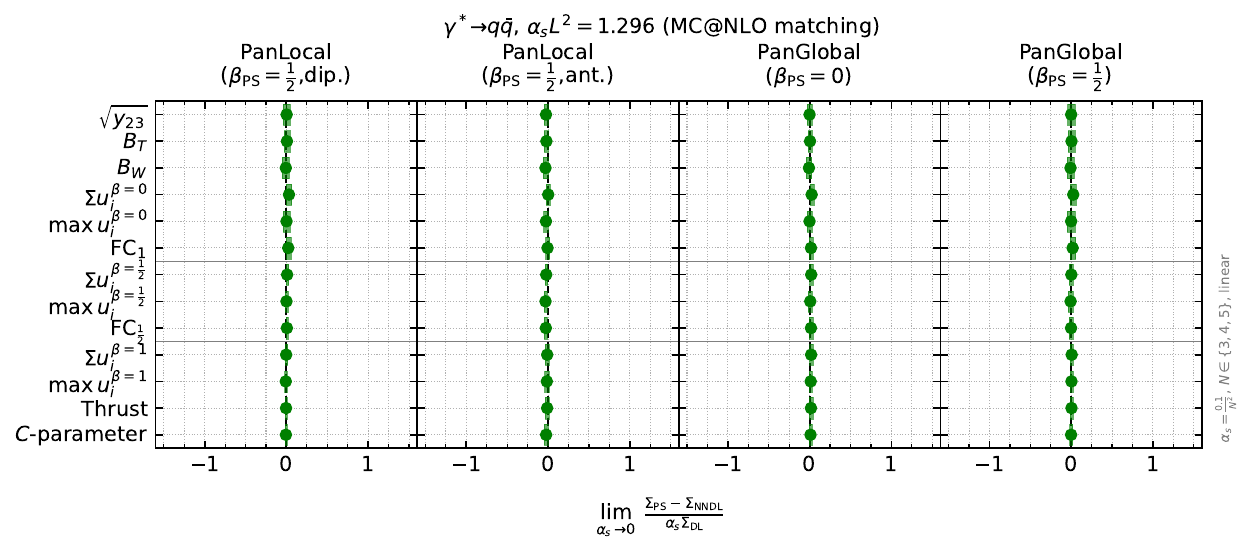}\\
  \includegraphics[width=.925\textwidth,page=22]{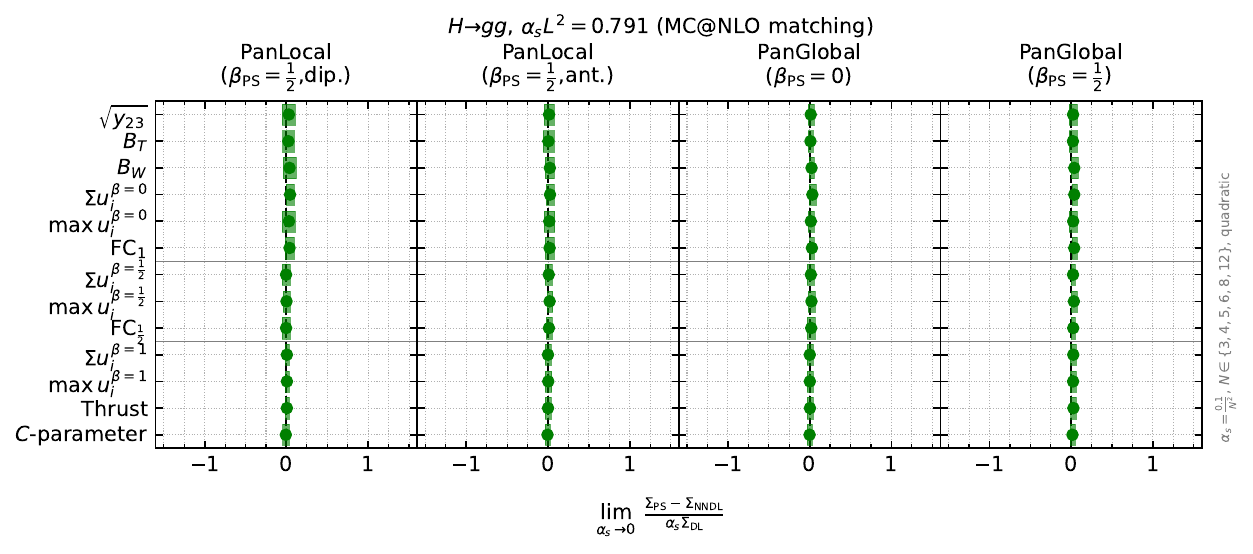}
  \caption{Results of the NNDL accuracy tests for the PanScales showers (see
    Fig.~\ref{fig:NNDL-no-matching}) matched with the \MCatNLO scheme.
  }
  \label{fig:NNDL-mcatnlo}
\end{figure}

In Fig.~\ref{fig:NNDL-mcatnlo}, we show results of the NNDL tests where the
PanScales showers are matched with the \MCatNLO scheme. Here as well, the
matched showers correctly reproduce all observables resummed to NNDL accuracy.

\begin{figure}[h]
  \centering
  \includegraphics[width=.925\textwidth,page=9]{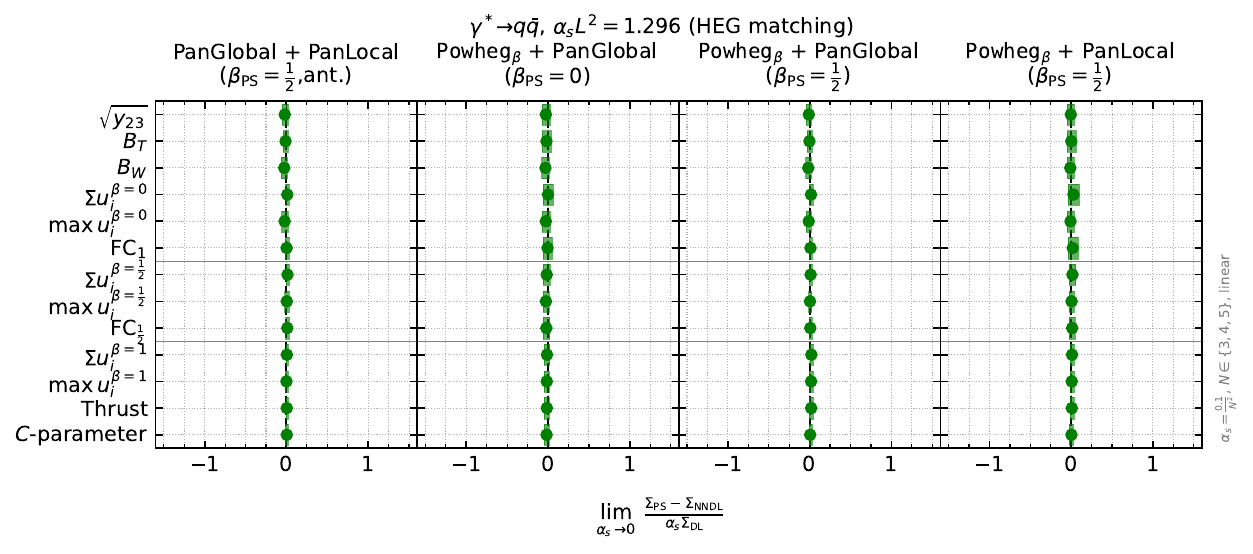}\\
  \includegraphics[width=.925\textwidth,page=21]{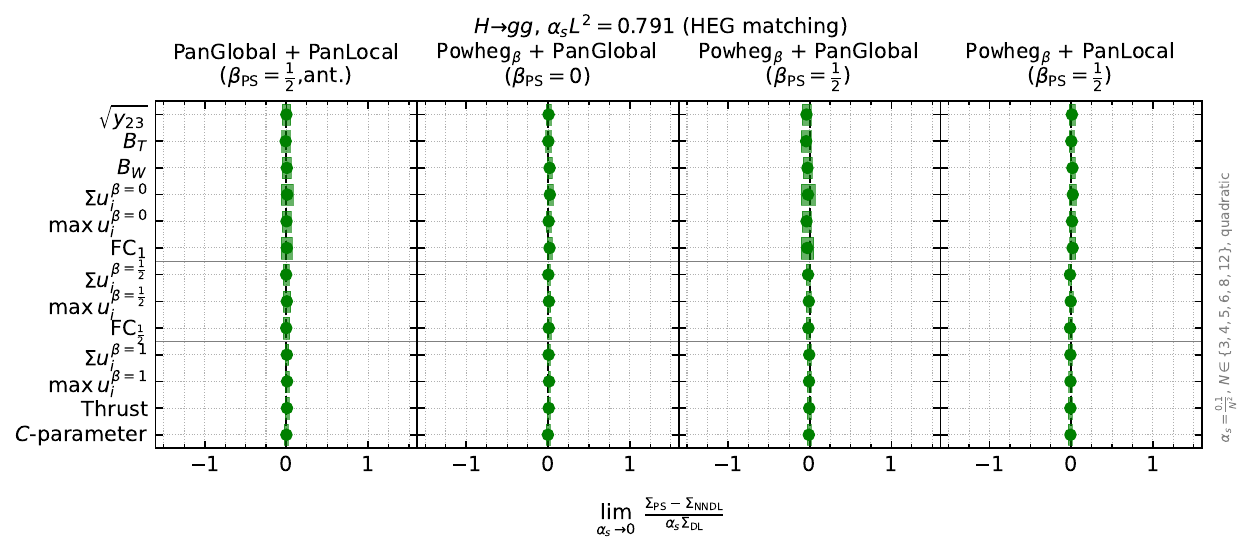}
  \caption{Results of NNDL accuracy tests for the four HEG/shower combinations
  shown in Table~\ref{tab:matching-combos}.}
  \label{fig:NNDL-powheg-2}
\end{figure}

We now turn to the case of \POWHEG matching. As summarised in
Table~\ref{tab:matching-combos}, we use either \POWHEG{$_\beta$} as a HEG, or
PanGlobal ($\betaps=\frac12$).
In Fig.~\ref{fig:NNDL-powheg-2}, four HEG/shower combinations are shown:
PanGlobal + PanLocal antenna ($\betaps=\frac12$), \POWHEG{$_\beta$} + PanLocal dipole
($\betaps=\frac12$) and \POWHEG{$_\beta$} + PanGlobal ($\betaps = 0,\frac12$). Where the
PanLocal (dipole or antenna) shower is used as the main shower, we veto
emissions according to the algorithm presented in section~\ref{sec:vetoing-alg}.
For all combinations presented in Fig.~\ref{fig:NNDL-powheg-2},
we also align the
choice of de-symmetrisation in the gluon splitting functions, $w^{\heg} =
w^{\ps} = 0$, following the analysis of section~\ref{sec:asym-mismatch}.
Results are all in agreement with NNDL.

\begin{figure}
  \centering
  \includegraphics[width=.495\textwidth,page=7]{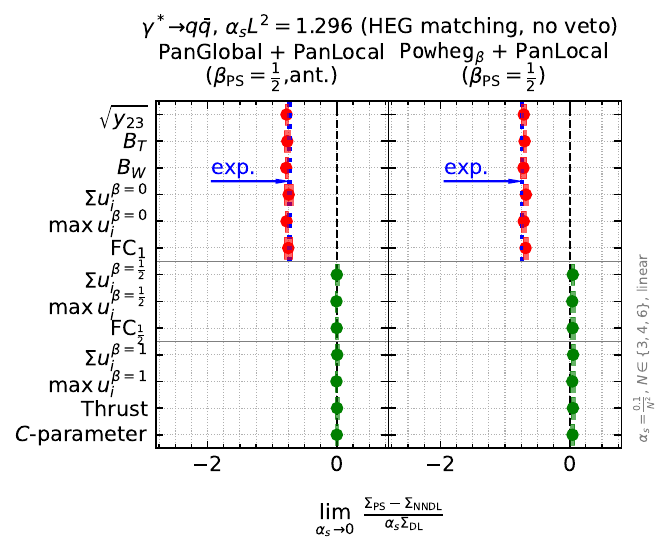}
  \includegraphics[width=.495\textwidth,page=19]{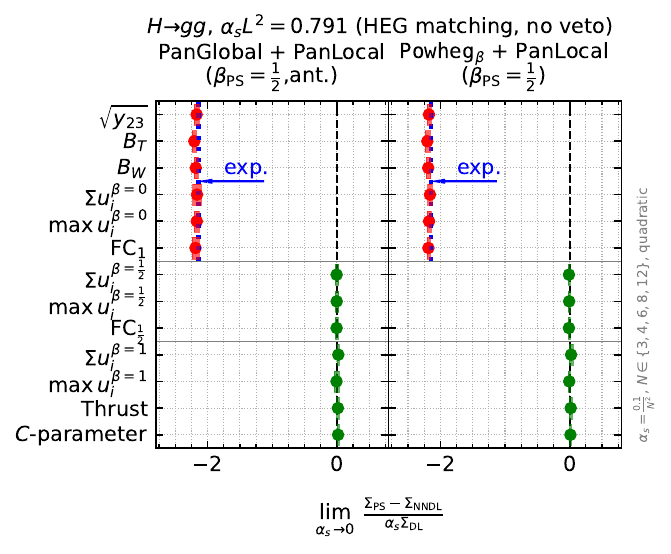}
  \caption{Results of NNDL accuracy tests for two combinations of HEG and shower
whose contours do not match in the hard-collinear region, all with $\betaps = \frac12$.
The showers are matched with the \POWHEG scheme, but the vetoing procedure given
in section~\ref{sec:vetoing-alg} is not applied, in order to highlight the NNDL
discrepancy expected for observables with $\betaobs<\betaps$. The expected value
from Eqs.~(\ref{eq:Delta-PG-PL})--(\ref{eq:Delta-PG-PL-h2gg}) is shown in dotted
blue (and marked by the blue arrow labelled ``exp.").} 
  \label{fig:NNDL-powheg-noveto}
\end{figure}
\begin{figure}
  \centering
  \includegraphics[width=.555\textwidth,page=24]{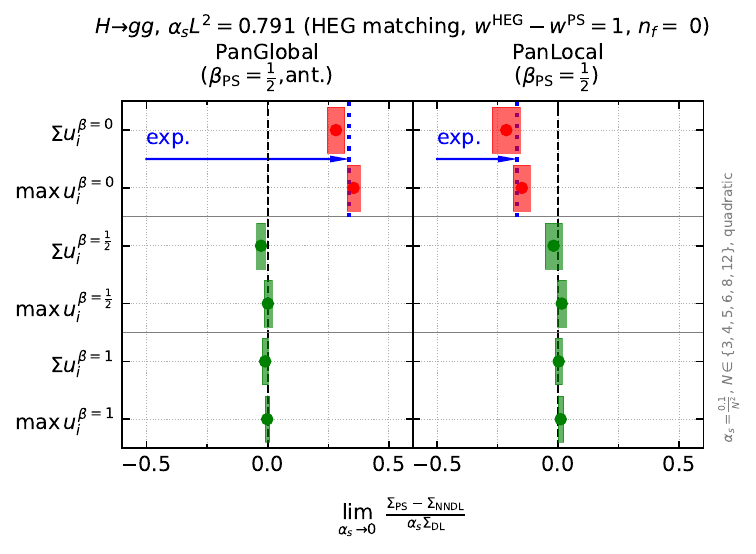}
  \caption{Results of NNDL accuracy tests with $n_f=0$ for the PanGlobal and PanLocal showers
    ($\betaps = \frac12$), using the same shower as both a HEG 
    and for the subsequent parton showering steps, but with different choices
    of the de-symmetrisation parameter in the gluon splitting function,
    $w^{\heg} = 1$, and $w^{\ps} = 0$ respectively.
    The expected value of the NNDL discrepancy for $\betaobs = 0$ from
    Eqs.~(\ref{eq:asym-mismatch-panglobal})--(\ref{eq:asym-mismatch-panlocal}) is
    shown in dotted blue.
  }
  \label{fig:NNDL-powheg-deltaw-1}
\end{figure}

Finally, we showcase the NNDL discrepancy arising from a failure to take the
considerations of section~\ref{sec:nndl-requirements} into account when matching
showers with the \POWHEG scheme.
In Fig.~\ref{fig:NNDL-powheg-noveto}, we display results of the NNDL tests for
two combinations of HEG/shower which require vetoing due to kinematic mismatch 
(PanGlobal + PanLocal antenna $\betaps=\frac{1}{2}$, and \powhegbeta + PanLocal
$\betaps=\frac{1}{2}$, see Table~\ref{tab:matching-combos}),
but where we deliberately do not apply the veto
algorithm of section~\ref{sec:vetoing-alg}. As anticipated, disabling the
veto has a visible effect for observables with $\betaobs < \betaps$. The
expected discrepancy for $\betaobs = 0$, which can be obtained by inserting
Eq.~(\ref{eq:Delta-PG-PL-2}) (for $\gamma^* \to q\bar q$) or
Eq.~(\ref{eq:Delta-PG-PL-h2gg}) (for $H \to gg$) into
Eq.~(\ref{eq:shower-HEG-line2}), is plotted
as a blue dotted line. That value is in agreement with the observed discrepancy
from the showers.

Similarly, we can investigate whether the numerical results confirm
the expected NNDL discrepancy stemming from the misaligned
de-symmetrisation of the gluon splitting functions presented in
section~\ref{sec:asym-mismatch}.
We show results of NNDL tests for a configuration where one of the PanGlobal or
PanLocal showers (with $\betaps = \frac12$), is used as a HEG, followed by the
same shower for subsequent emissions. In order to see the discrepancy of
section~\ref{sec:asym-mismatch}, we choose different values of the
de-symmetrisation parameter $w_{gg}$, see
Eq.~(\ref{eq:pgg-asym}), for the first emission ($w^{\heg}$) and
for the rest of the showering ($w^{\ps}$).
As can be seen from Eqs.~(\ref{eq:asym-mismatch-panglobal})
and~(\ref{eq:asym-mismatch-panlocal}), the discrepancy in both cases is
proportional to $C_A - n_f T_R$.
In order to avoid the large cancellation of this effect with
$n_f = 5$, in Fig.~\ref{fig:NNDL-powheg-deltaw-1} we run with
$n_f = 0$, and we set $w^{\heg} = 1$ and $w^{\ps} = 0$.
While we could have extracted the $C_1$ coefficients numerically for
$n_f=0$ as well, in Fig.~\ref{fig:NNDL-powheg-deltaw-1} we only show
observables for which the analytic form of $C_1$ is available.
Recall that we only expect a discrepancy for $\betaobs < \betaps$.
Though the NNDL discrepancy associated with mismatched
$w^{\heg} \neq w^{\ps}$ is numerically smaller than for the case of
the kinematic mismatch (note the different scale on the $x$-axis of
Fig.~\ref{fig:NNDL-powheg-deltaw-1}), we find that the results agree
with our analytic predictions for them, both when there are
discrepancies and when there are none.

\section{Phenomenological considerations}
\label{sec:pheno}

In this last section, we briefly explore the interplay of matching and
logarithmic accuracy with physical choices for the coupling
and values of observables, as opposed to the asymptotic values used in
the preceding sections.
The intent at this stage is not to be exhaustive, nor to compare to
data (for that we would still like to have finite quark mass effects
and an interface to hadronisation), but rather to get some insight
into how logarithmic-accuracy improvements affect practical
distributions.

We will show parton-level results with a preliminary estimate of
uncertainties, specifically taking an envelope of two sources of
uncertainty: (1) renormalisation scale variation and (2) uncertainties
associated with residual lack of control of shower matrix elements
beyond the matched emission.

The scale uncertainties are calculated according to
Ref.~\cite{vanBeekveld:2022ukn}'s adaptation of the prescription by
Mrenna and Skands~\cite{Mrenna:2016sih}.
Specifically, for showers that are NLL accurate, we take the emission
intensity to be proportional to
\begin{equation}
  \label{eq:xmuR}
  \as(\muR^2)\left(1 + \frac{K \as(\muR^2)}{2\pi}
    + 2 \as(\muR^2) b_0 (1-z) \ln \xR \right)\,,
  \qquad
  \muR = \xR \muR^{\text{central}}\,.
\end{equation}
Here $z$ is the fraction of the emitter momentum carried away by the
radiation,\footnote{Specifically, in \powhegbeta we take $z$ 
  equal to $\xi$ in Eq.~(\ref{eq:powheg-variables}).
  For the PanGlobal and PanLocal showers, we take it equal to
  $a_k$ and $b_k$ in Eq.~(4) of Ref.~\cite{Dasgupta:2020fwr},
  respectively, when generating the $g(\bar \eta)$ and $g(-\bar \eta)$
  terms.}
while $b_0$ and $K$ are the usual $\beta$-function and
CMW~\cite{Catani:1990rr} coefficients, and $\muR^{\text{central}}$ is
the emission transverse momentum ($\kappa_\perp$) as defined in the shower.
The factor of $1-z$ ensures that NLO scale compensation is present for
soft-collinear emissions, but turned off for hard emissions.\footnote{One might
argue that scale compensation should be completely turned off
earlier, e.g. for $z > \frac12$.
We leave exploration of different possible schemes to future work.}
Scale variation will be probed by taking $\xR = \{\frac12,1,2\}$.
We will use this scale variation also in the matching, e.g.\ so that
HEG-style matching has the correct scale compensation in the
infrared.
Throughout, we use the NODS colour scheme~\cite{Hamilton:2020rcu}
(i.e.\ full-colour NLL for global event shapes), a
two-loop coupling, with $\as(m_Z^2) = 0.118$, $5$ light flavours and
an infrared cutoff implemented such that $\as(\muR^2)$ is set to zero
for $\muR < \xR \times 0.5\GeV$.

We will also show results with our PanScales
implementation~\cite{Dasgupta:2020fwr} of the \pythiaeight\
shower~\cite{Sjostrand:2004ef} (which we call \pspythiaeight).
Since it is a LL shower, we will not include the scale compensating
terms in Eq.~(\ref{eq:xmuR}) for shower emissions (nor for the HEG
emission), however we do include a two-loop running and the CMW constant term
$K$.

To estimate the uncertainty associated with lack of control of shower
matrix-elements in the hard region, we modify the default shower
splitting probability according to 
\begin{equation}
  \label{eq:xhard}
  P_\text{splitting}(\xhard) =
  P_\text{splitting}^\text{(default)} \times
  \left[1
    + (\xhard-1) \min\left(\frac{4\kappa_\perp^2}{Q^2}, 1\right)
  \right]\,,
\end{equation}
where $\xhard=1$ reproduces the default splitting probability and we
take as variations $\xhard = \{\frac12,1,2\}$.
As with the scale variation, there is some arbitrariness in these
choices and their detailed implementation, whose full investigation we
leave to future work together with that of other potential sources of
uncertainty.
For unmatched shower results, we apply Eq.~(\ref{eq:xhard}) to all
splittings, while for matched shower results we apply it to all but
the first emission.\footnote{In the \MCatNLO procedure, $\xhard$
  variation should ideally be done at all stages of the shower, with a
  compensating $\xhard$ dependence in the $(R-\Rps)$ term of
  Eq.~(\ref{eq:MCatNLO}).
  We have not yet implemented this, and accordingly we do not perform
  the $\xhard$ variation for the \MCatNLO runs.
}
The results are shown without spin correlations, but we have verified
for the multiplicative matching procedure that they have a numerically
negligible impact on the event-shape type observables shown here (and
no impact up to NLL/NNDL).
\logbook{7cfa534f3}{See 2020-eeshower/analyses/global-obs-pheno/matching-paper-runs-GPS-19122022/spin-check.pdf}%
Note that fixed-order tests of spin correlations with matching are
given in Appendix~\ref{sec:spin}. 

\begin{figure}
  \centering
  \includegraphics[width=0.33\textwidth,page=7]{figures/3jet-distribution-examples.pdf}%
  \includegraphics[width=0.33\textwidth,page=8]{figures/3jet-distribution-examples.pdf}%
  \includegraphics[width=0.33\textwidth,page=9]{figures/3jet-distribution-examples.pdf}%
  \\
  \includegraphics[width=0.33\textwidth,page=1]{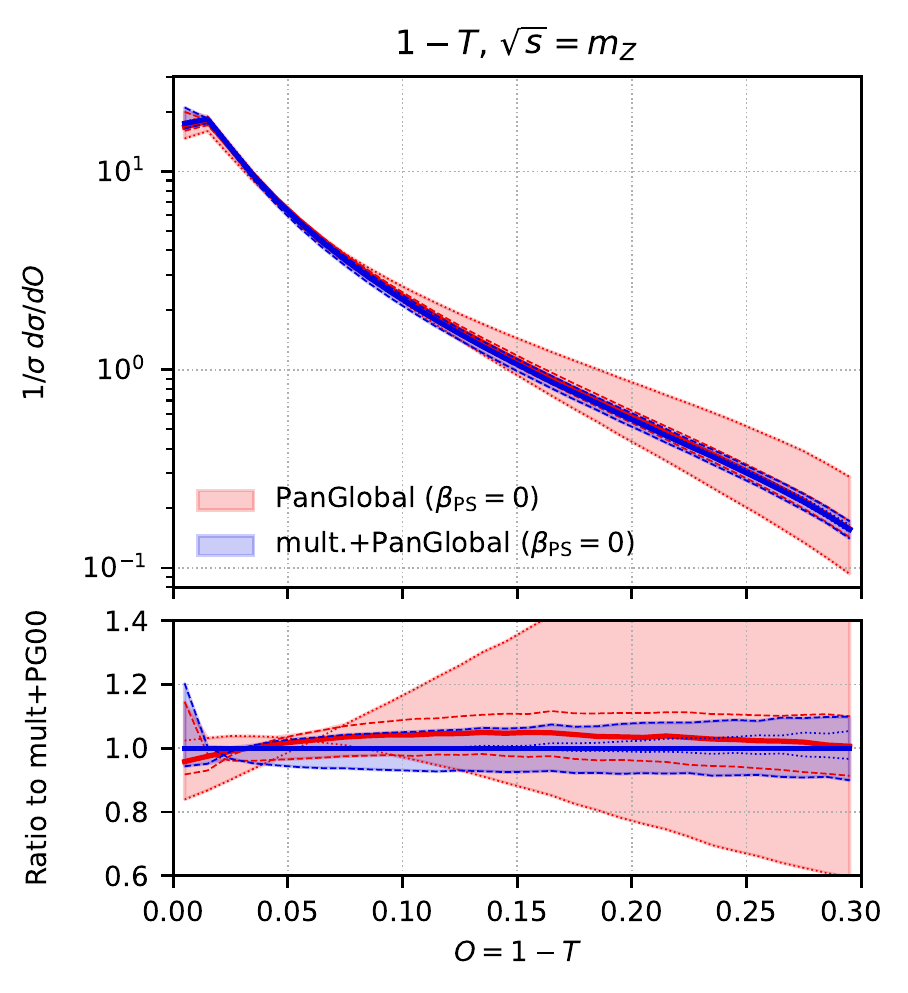}%
  \includegraphics[width=0.33\textwidth,page=2]{figures/3jet-distribution-examples.pdf}%
  \includegraphics[width=0.33\textwidth,page=3]{figures/3jet-distribution-examples.pdf}%
  \\
  \includegraphics[width=0.33\textwidth,page=4]{figures/3jet-distribution-examples.pdf}%
  \includegraphics[width=0.33\textwidth,page=5]{figures/3jet-distribution-examples.pdf}%
  \includegraphics[width=0.33\textwidth,page=6]{figures/3jet-distribution-examples.pdf}%
  \caption{Thrust (left), Cambridge $\ln y_{23}$ (middle) and SoftDrop
    $\ln k_t/Q$ (right) distributions, unmatched (red) and matched (blue).
    They are obtained with a LL shower (our PanScales implementation of the
    \pythiaeight shower (\pspythiaeight, top row)) and
    two NLL showers: PanGlobal with 
    $\betaps = 0$ (middle row) and PanLocal $\betaps=\frac12$ (bottom
    row). 
    The last row also shows the impact of HEG-style matching without
    the veto discussed in section~\ref{sec:vetoing-alg}.
    Dotted lines show $\xhard$ variation, while dashed lines show
    $\xR$ variations.
  }
  \label{fig:pheno-plots-thrust}
\end{figure}

Fig.~\ref{fig:pheno-plots-thrust} shows parton-level results for the
thrust and Cambridge $\ln y_{23}$ at $\sqrt{s} = m_Z$.
It also features a SoftDrop (SD) $\ln k_t/Q$ distribution, with
$z_\text{cut}= 0.25$,
$\beta_\text{SD} = 0.0$~\cite{Dasgupta:2013ihk,Larkoski:2014wba} and
$k_t$ for a $\itilde \to ik$ splitting defined as
$\min(E_i,E_k)\sin\theta_{ik}$.
The choice of $z_\text{cut}= 0.25$ is larger than commonly used, and
helps concentrate on the hard-collinear region. 
The SoftDrop procedure is applied to each of the two jets as obtained
from a $2$-jet Cambridge clustering.
This observable is shown for $\sqrt{s} = 2\TeV$ insofar as it is
intended to be illustrative of an LHC jet substructure observable.
%

%
The top row of Fig.~\ref{fig:pheno-plots-thrust} shows results for our
implementation of the \pythiaeight shower. 
Recall that since this shower is LL rather than NLL we do not include
the scale compensation terms of Eq.~(\ref{eq:xmuR}) when varying the
renormalisation scale (neither in the shower itself, nor in the
\powhegbeta stage).\footnote{We include a kinematic veto in the hard
  collinear region, since \powhegbeta and \pythiaeight kinematics do not
  match up there.
  However, the fact that we start the shower from $v^\ps = v^\heg$
  means that we do not address under-counting in the soft regions at
  angles that bisect the dipoles in the centre-of-mass frame.
  Ultimately it would be of interest to replicate exactly what is done
  in standard {\POWHEG}+\pythiaeight usage.
  From a logarithmic point of view, the lack of NLL accuracy in
  \pythiaeight would anyway prevent this combination from achieving
  NNDL accuracy.
  \logbook{272ac426d7d24}{We explored setting $v^\ps = Q$. This alone is not enough
    because we require $v^\ps<v^\heg$ for non-Born emitters, whereas
    we should really generalise our veto to also handle those non-Born
    emitters. Still it gives an indication of the potential size of
    residual effects, and these were at the $2\%$ level.
    Plots with a starting $v^\ps = Q$ (``power shower'') are to be
    found in
    2020-eeshower/analyses/global-obs-pheno/matching-paper-runs-LS-17012023-mZ/3jet-column-summary.withpowershower.pdf
    and
    2020-eeshower/analyses/global-obs-pheno/matching-paper-runs-LS-17012023-mZ/3jet-column-summary.wopowershower.pdf
    One might reasonably expect the $2\%$ to double if we do
    everything correctly.
    A further remark is that the \POWHEG map for FSR looks different
    from ours even for $\betaps=0$, in particular in the soft region
    it uses constant $E_k^2(1-\cos\theta)$, whereas we have constant
    $E_k^2 \sin^2\theta$.
    The true \POWHEG map looks more like the Pythia8 map, but note that
    the $k_t$ ``dip'' that is at central rapidity for the first step
    is instead at soft-collinear rapidities (CoM dipole midpoint)
    in the next step --- so it's far from clear to figure out how to
    get everything to fit together...
  }
}
The remaining rows show the PanGlobal shower with $\betaps = 0$
and the PanLocal (dipole) shower $\betaps=\frac12$.
Without matching, there are large uncertainties in the $3$-jet region,
mostly dominated by the $\xhard$ variations (dotted
lines).\footnote{For the PanGlobal and \pspythiaeight showers the
  $\xhard$ variation is a reasonable reflection of the uncertainties
  in that region.
  For the PanLocal shower, it seems to underestimate the
  uncertainties.
  One might consider extending the lower $\xhard$ limit to $0$ rather
  than $\frac12$, but we leave further study of this question to
  future work.  }
Once matching is turned on, it is the scale uncertainties that
dominate the uncertainty bands over the full range shown for the event
shapes.
Note that the scale compensation that is used in the NLL shower brings a
visible reduction in uncertainty as compared to the case of LL
showers. 

Observe also that \powhegbeta matching with the PanLocal shower
without the kinematic veto, shown in the lower row (green band),
induces a noticeable shift in the kinematic distributions.
Curiously, this is true not just for the $\ln y_{23}$ distribution
(where there is a NNDL effect), but also for the thrust (where there
is no NNDL effect, but still a kinematic double-counting).
The clearest effect is seen for the SoftDrop observable at
intermediate values of $\ln k_t/Q$.
This is not unexpected: recall from the discussion at the end of
section~\ref{sec:kinematic-mismatch} that for moderately large values
of the logarithm, $L \sim 1/\sqrt{\as}$, the impact of the HEG/shower
contour mismatch is expected to be of relative order $\sqrt{\as}$
rather than at most $\as$ for standard global event shapes.

\begin{figure}
  \centering
  \includegraphics[width=0.99\textwidth,page=3]{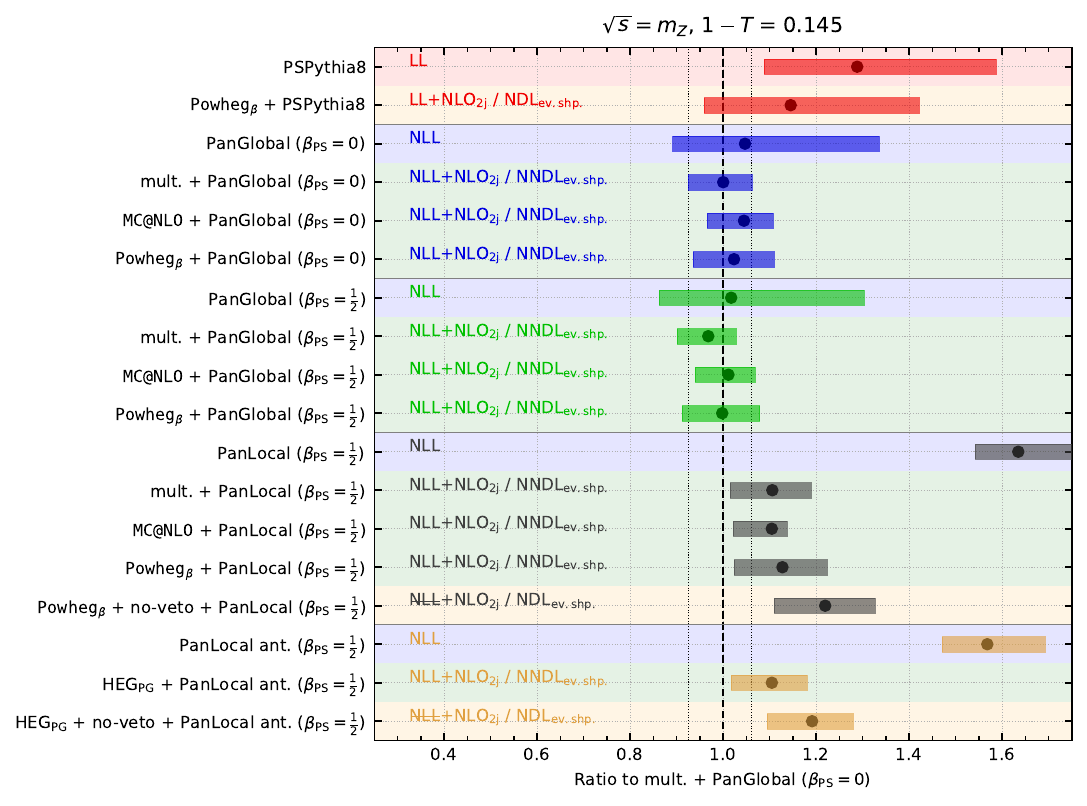}
  \caption{Ratio of multiple shower/matching scheme combinations
  to the multiplicatively-matched PanGlobal $\betaps = 0$ shower
  in a bin $0.14 < 1-T < 0.15$ of the thrust distribution.
  The error bands represent the scale uncertainty and are colour-coded
  differently for each ``main" shower, while the background colour
  reflects the accuracy of the matching/shower combination, i.e.\ LL
  (red), LL+NLO$_{2j}$/NDL$_\mathrm{ev. shp.}$ or
  \sout{NLL}+NLO$_{2j}$/NDL$_\mathrm{ev. shp.}$ (yellow), NLL 
  (blue), and NLL+NLO$_{2j}$/NNDL$_\mathrm{ev. shp.}$ (green).
  The thin dotted vertical lines indicate the size of the scale
  uncertainties for the reference results, i.e.\
  multiplicatively-matched PanGlobal $\betaps=0$.
  Recall from section~\ref{sec:kinematic-mismatch} that the \sout{NLL}
  notation indicates a breaking of exponentiation. }
  \label{fig:pheno-plots-thrust-bin}
\end{figure}

\begin{figure}
  \centering
  \includegraphics[width=0.99\textwidth,page=8]{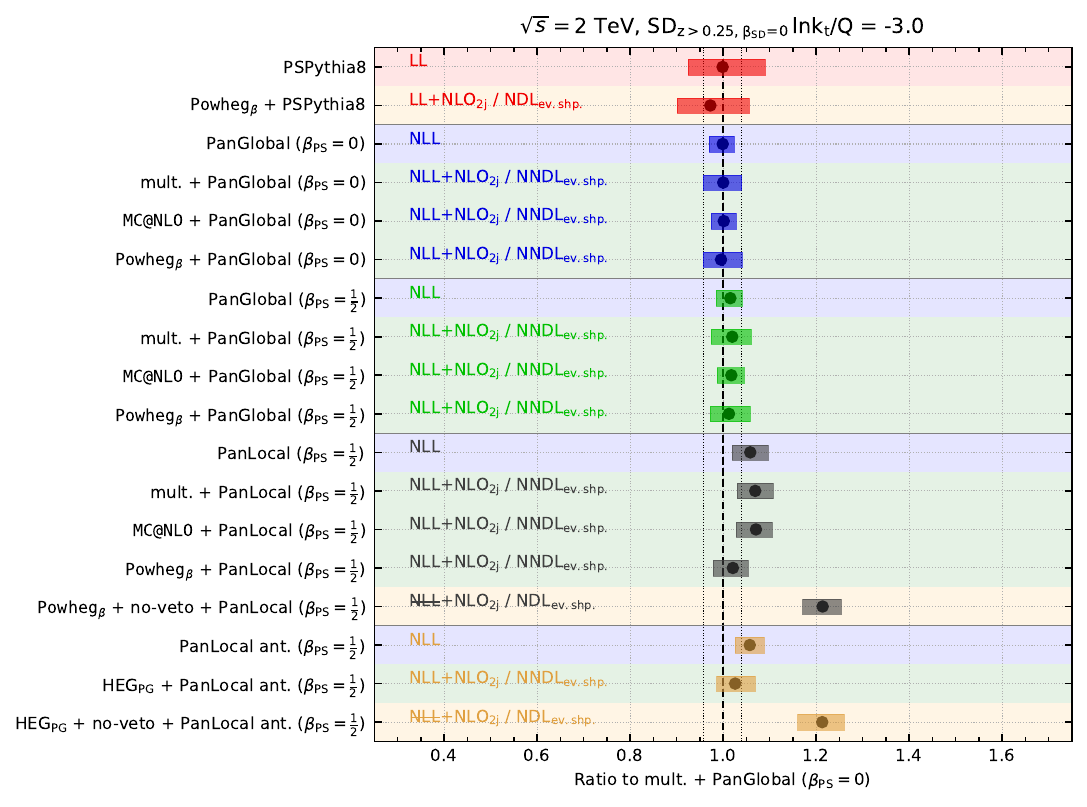}
  \caption{Analogue of Fig.~\ref{fig:pheno-plots-thrust-bin} for the
    $k_t/Q$ distribution of the splitting from the SoftDrop
    ($z_\text{cut}=0.25, \beta_\text{SD}=0$) procedure, in a bin
    $-3.1<\ln k_t/Q<-3.0$.
  }
  \label{fig:pheno-plots-SD-bin}
\end{figure}

In Fig.~\ref{fig:pheno-plots-thrust} it was possible to show only a
limited number of matching/shower combinations.
To help visualise characteristics of a wider range shower/matching
combinations,
we select a specific bin in each of two distributions,
$0.14 < 1-T < 0.15$, and the SoftDrop $-3.1<\ln k_t/Q<3.0$, chosen so
as to be in a transition region between the large-logarithm and hard
$3$-jet regimes.
We then examine the ratio of a range of showers and matching schemes
to the multiplicatively-matched PanGlobal $\betaps = 0$ result.
This is shown in Figs.~\ref{fig:pheno-plots-thrust-bin} and
\ref{fig:pheno-plots-SD-bin}, with one row for each shower/matching
combination.

Features to note beyond those observed for
Fig.~\ref{fig:pheno-plots-thrust} are:
(1) The matched NLL showers have uncertainties that are broadly
similar across matching and shower combinations (somewhat larger for
PanLocal $\betaps=\frac12$), and consistent with each other to within
uncertainties.
Note that the residual $10\%$ variation between matched showers could
have a significant impact on tuning, and ideally one should include
$3$-jet NLO matching, e.g.\ as done in Ref.~\cite{Hartgring:2013jma}.
(2) The characteristics seen in Fig.~\ref{fig:pheno-plots-thrust} for
\powhegbeta matching with PanLocal (dipole) $\betaps=\frac12$ without
the kinematic veto appear to be replicated also when using the
PanGlobal shower as a HEG (HEG$_\text{PG}$) in combination with the PanLocal (antenna)
shower without a veto, suggesting that they are genuinely an impact of
the lack of veto rather than a coincidence. 

Overall the results in this section highlight the importance of both
matching and NLL accuracy and of bringing them together consistently,
lending support to the practical value of pursuing the programme to
improve shower logarithmic accuracy.

\section{Conclusions}
\label{sec:conclusions}

The simple framework of two-body decays that we have studied here
provides a clean and powerful laboratory to explore the interplay of
parton-shower logarithmic accuracy and matching.
In particular, we have seen that NLO matching can augment the accuracy
of NLL showers, so that they additionally attain NNDL accuracy for
global event shapes.
This was relatively straightforward to achieve with multiplicative and
\MCatNLO matching methods, because they alter the shower behaviour or
add events only in the hard region.
In contrast, with matching methods such as \POWHEG that take
responsibility for generating the hardest emission, an extra element is
needed, which is to ensure that the hardest-emission generator and
shower align in their generation of phase space in the full soft
and/or collinear regions.
Failing to account for this prevents the HEG/shower combination from
attaining NNDL accuracy.
Furthermore, it subtly compromises NLL accuracy, generating spurious
super-leading logarithms, Eq.~(\ref{eq:explicit-lnSigma}), that resum
in such a way, Eq.~(\ref{eq:shower-HEG-line2}), as to vanish in
standard numerical NLL accuracy global event-shape tests (but not
necessarily for single logarithmic observables, such as SoftDrop with
$\beta_\text{SD}=0$).
In this paper we used the (standard) approach of vetoing shower steps
in order to avoid double-counting phase space already generated with
the HEG.
However, thinking forward to possible approaches to achieving yet
higher logarithmic accuracy, it is likely to be advantageous to
consider designing HEG tools such that they have the freedom to mimic
the lowest order soft/collinear phase-space generation of any given
shower.

A related and more subtle issue occurs when a given phase-space point
can be reached from more than one value of the HEG or shower ordering
variable.
In our study, this issue arose in the context of de-symmetrisation of
gluon splitting functions in the hard-collinear region, cf.\
section~\ref{sec:asym-mismatch}. 
However, we expect it to be relevant more generally also in processes
with non-trivial soft large-angle structures, for example in the
presence of three or more Born partonic legs.
The critical observation is that in such situations, the HEG and the
shower must have the same relative weights for each of the distinct
values of the ordering variable that can lead to that phase-space
point.

The numerical tests of section~\ref{sec:log-tests} provided extensive
validation of our understanding of parton-shower NNDL event-shape
accuracy.
The results confirmed NNDL accuracy in the situations where we expected
to achieve it, and furthermore reproduced the analytic expectations for
discrepancies in situations with kinematic or de-symmetrisation
mismatches.

In section~\ref{sec:pheno}, we took first steps towards the
exploration of the phenomenological impact of logarithmically accurate
showers, including preliminary uncertainty estimates.
Perhaps the main conclusion to be drawn so far is that there is
good consistency across different matching schemes and showers, to
within the uncertainties, as well as significant reductions in
uncertainties relative to LL showers.
Furthermore, matching/shower combinations that do not achieve NNDL
accuracy appear to give predictions in tension with those from
NNDL-accurate showers.
Clearly next important phenomenological steps include interfacing with
hadronisation, the inclusion of heavy quark masses, comparisons to
data and associated exploration of tuning NLL showers.

\section*{Acknowledgements}

We are grateful to our PanScales collaborators (Melissa van Beekveld,
Mrinal Dasgupta, Fr\'ed\'eric Dreyer, Basem El-Menoufi, Silvia Ferrario
Ravasio, Jack Helliwell, Rok Medves, Pier Monni, Gr\'egory Soyez and
Alba Soto Ontoso), for their work on the code, the underlying
philosophy of the approach and comments on this manuscript.
We also wish to thank Paolo Nason for discussion about the
manuscript. 
This work was supported
by a Royal Society Research Professorship
(RP$\backslash$R1$\backslash$180112) (GPS, LS),
by the European Research Council (ERC) under the European Union’s
Horizon 2020 research and innovation programme (grant agreement No.\
788223, PanScales) (KH, AK, GPS, LS, RV), 
by the Science and Technology Facilities Council (STFC) under
grants ST/T000856/1 (KH) and ST/T000864/1 (GPS)
and by Somerville College (LS) and Linacre College (AK).
GPS, LS, and RV would
like to thank the CERN Theory Department for hospitality while part of
this work was carried out.
\appendix

\section{$C_1$ coefficients}
\label{sec:appC1}

In this Appendix, we report the numerical values for the $C_1$
coefficients that enter into the NNDL resummation. These can be found
in Table~\ref{tab:c1-coeff}, alongside analytical expressions where
these can be found in literature. The numerical values are quoted for
$C_A = 3$, $C_F=\frac{4}{3}$ and $n_f = 5$ and the statistical error
on the last digit
is reported in the brackets.

The numerical extraction of $C_1$ is performed with our own
  matrix-element integrator.
It uses the phase space of Ref.~\cite{Dokshitzer:1992ip} for
the $\gamma\rightarrow q\bar q$ channel and the phase space
of Ref.~\cite{Cox:2018wce} for the $H\rightarrow gg$ channel which is
better suited due to the symmetric nature of the $H\rightarrow ggg$
process. The relevant matrix elements are given by
\begin{align}
  \frac{|\mathcal{M}_{\gamma\rightarrow q(1)\bar q(2) g(3)}|^2}{\sigma_0} &= \frac{\as C_F}{2\pi}\frac{x_1^2 + x_2^2}{(1-x_1)(1-x_2)}, \\
  \frac{|\mathcal{M}_{h\rightarrow g(1) g(2) g(3)}|^2}{\sigma_0} &= \frac{\as C_A}{6\pi}\frac{1 + (1-x_1)^4 + (1-x_2)^4 + (1-x_3)^4}{(1-x_1)(1-x_2)(1-x_3)}, \\
  \frac{|\mathcal{M}_{h\rightarrow g(1) q(2) \bar q(3)}|^2}{\sigma_0} &= \frac{\as n_f T_R}{\pi}\frac{(1-x_2)^2 + (1-x_3)^2}{1-x_1},
\end{align}
where $x_i$ is the energy fraction of the $i$-th particle, and
$\sigma_0$ refers to the underlying Born cross section for the
processes $\gamma\rightarrow q\bar q$ and $H\rightarrow gg$
respectively. In the limit where any of the global event shapes that
we study in this paper go singular the cumulative distribution takes
the form of Eq.~\eqref{eq:global-evs-resum-intro}, and at fixed first
order can be written in the form~\cite{Catani:1991kz}
\begin{equation}
\Sigma(O < e^L) =
    1 + \frac{\as}{2\pi} \, [\, G_{12} \, L^2 + G_{11} \, L + C_1 \,] \, ,
\label{eq:global-evs-FO}
\end{equation}
where the constants $G_{mn}$ 
depend only on the infrared and collinear scaling of the global event
shape, and can be extracted from CAESAR~\cite{Banfi:2004yd}. In
order to extract $C_1$ it is therefore enough to compute numerically
the normalised cumulative distribution in the singular region, subtracting the two
logarithmic terms analytically. In practice we go to values of
$\ln\mathcal{O}\sim -16$ and have excellent agreement with the known
coefficients, where available.

\begin{landscape}
\begin{table}[t] 
  \centering
  \phantom{x}\medskip
  \normalsize{
  \begin{tabular}{l|ccc|ccc}
    \toprule
    Observable & \multicolumn{3}{c|}{$\gamma^* \to q\bar{q}$ } &  \multicolumn{3}{c}{$H \to gg$}\\
    \midrule
    & Numerical &  \multicolumn{2}{c|}{Analytic} & Numerical &  \multicolumn{2}{c}{Analytic} \\
    \midrule
    $\sqrt{y_{23}}$                 & -6.685(2)   & $C_F\left(-6\ln 2 + \frac{\pi^2}{6}-\frac{5}{2}\right)$  & \cite{Banfi:2001bz}  & -13.417(6) &  &  \\
    $B_T$                           &  1.825(2)   & $C_F\left( \pi^2-\frac{17}{2} \right)$                       & \cite{Becher:2012qc}  & 6.438(5)   &  &  \\
    $B_W$                           &  1.825(2)   & $C_F\left( \pi^2-\frac{17}{2} \right)$                       & \cite{Becher:2012qc}  & 6.438(5)   &  &  \\
    $\Sigma\,u_i^{\beta=0.0}$       & -4.492(2)   & $C_F \left(-6\ln 2+\frac{\pi^2}{3}-\frac{5}{2}\right)$   & \cite{Medves:2022ccw}  & -8.482(5)  & $C_A\left(-\frac{22}{3} \ln 2+\frac{\pi^2}{3} -\frac{44}{9}\right) + n_f \left(\frac{4}{3}\ln 2 + \frac{25}{18}\right)$& \cite{Medves:2022ccw} \\
    $\mathrm{max}\,u_i^{\beta=0.0}$ & -4.492(2)   & $C_F \left(-6\ln 2+\frac{\pi^2}{3}-\frac{5}{2}\right)$   & \cite{Medves:2022ccw}  & -8.482(5)  & $C_A\left(-\frac{22}{3} \ln 2+\frac{\pi^2}{3} -\frac{44}{9}\right) + n_f \left(\frac{4}{3}\ln 2 + \frac{25}{18}\right)$& \cite{Medves:2022ccw} \\
    ${\rm FC}_{1}$                  &  1.825(2)   &                                                              &   & 6.439(5)   &  &  \\
    \midrule
    $\Sigma\,u_i^{\beta=0.5}$       &-2.9013(16)  &$C_F\left( -4\ln 2 + \frac{\pi^2}{9} - \frac{1}{2}\right)$ & \cite{Medves:2022ccw}  & -6.595(4)  & $C_A\left(-\frac{44}{9} \ln 2+\frac{\pi^2}{9} - \frac{127}{54}\right) + n_f \left(\frac{8}{9}\ln 2 + \frac{23}{27}\right)$ & \cite{Medves:2022ccw} \\
    $\mathrm{max}\,u_i^{\beta=0.5}$ &-2.9013(16)  &$C_F\left( -4\ln 2 + \frac{\pi^2}{9} - \frac{1}{2}\right)$ & \cite{Medves:2022ccw}  & -6.595(4)  & $C_A\left(-\frac{44}{9} \ln 2+\frac{\pi^2}{9} - \frac{127}{54}\right) + n_f \left(\frac{8}{9}\ln 2 + \frac{23}{27}\right)$ & \cite{Medves:2022ccw} \\
    ${\rm FC}_{\frac{1}{2}}$        & 1.3106(16)  &                                                              &   & 3.352(3)   &  &  \\
    \midrule
    $\Sigma\,u_i^{\beta=1.0}$       & -2.1066(13) & $C_F \left(-3 \ln 2 + \frac{1}{2}\right)$                & \cite{Medves:2022ccw}  & -5.650(3)  & $C_A\left(-\frac{11}{3} \ln 2-\frac{13}{12}\right) + n_f \left(\frac{2}{3} \ln 2 + \frac{7}{12}\right)$ & \cite{Medves:2022ccw} \\
    $\mathrm{max}\,u_i^{\beta=1.0}$ & -2.1066(13) & $C_F \left(-3 \ln 2 + \frac{1}{2}\right)$                & \cite{Medves:2022ccw}  & -5.650(3)  & $C_A\left(-\frac{11}{3} \ln 2-\frac{13}{12}\right) + n_f \left(\frac{2}{3} \ln 2 + \frac{7}{12}\right)$ & \cite{Medves:2022ccw} \\
    $1-T$                           & 1.0522(13)  & $C_F\left( \frac{\pi^2}{3}-\frac{5}{2} \right)$              & \cite{Catani:1992ua}   &  1.810(3)  &  &  \\
    $C$                             & 5.4381(12)  & $C_F\left( \frac{2\pi^2}{3}-\frac{5}{2}\right)$              & \cite{Catani:1998sf}  &  11.680(3) &  &  \\
    \bottomrule
  \end{tabular}}
  \caption{$C_1$ coefficients for the observables under scrutiny in
    the NNDL tests of section~\ref{sec:log-tests}.
    The numerical
    values are extracted with $C_A = 3$, $C_F=\frac{4}{3}$ and $n_f =
    5$.
    The observables are organised into three groups with $\betaobs$
    values respectively of $0$, $\frac12$ and $1$.
    \logbook{}{see c1-coefficients.nb for numerical values from
      formulas in order to check agreement; only the $C$-parameter
      differs by more than $1\sigma$ (about $1.3\sigma$)}
  } 
\label{tab:c1-coeff}
\end{table}
\end{landscape}

\section{Spin correlations in the matching region}
\label{sec:spin}

The inclusion of spin correlation effects~\cite{Richardson:2001df,Fischer:2017htu,Richardson:2018pvo} constitutes a defining component of
a NLL-accurate parton shower. Purely collinear \cite{Karlberg:2021kwr} and soft
\cite{Hamilton:2021dyz} spin correlations were previously included in the
PanScales showers, and it was shown that the expected logarithmic structure is
reproduced. In this appendix, we briefly validate the extension of the algorithm
to account for matching.
To the best of our knowledge, analytical and semi-numerical
resummation results for spin-sensitive observables such as the triple
energy-correlation~\cite{Chen:2020adz} and Lund-declustering azimuthal
correlations~\cite{Karlberg:2021kwr} have yet to
be extended to include $3$-jet matching.
We thus
validate the algorithm at fixed-order, where focusing on the matching
region is straightforward.

The extension of the Collins-Knowles spin algorithm to account for matched
branchings is straightforward (and has been explored already in
Ref.~\cite{Richardson:2018pvo}).
As formulated in
Ref.~\cite{Hamilton:2021dyz}, the 
algorithm relies on a two-body hard scattering amplitude
$\mathcal{M}_{\text{hard}}^{\lambda_1 \lambda_2}$ that is positioned at the root
of the binary tree structure that is used to efficiently transmit spin
information. Normally, when a shower branching occurs, a corresponding node is
attached to this binary tree that encodes the spin information of the
branching. However, when the first branching is matched, one should instead
re-initialise the binary tree with the corresponding three-body hard scattering
amplitude $\mathcal{M}_{\text{hard}}^{\lambda_1 \lambda_2 \lambda_3}$.
Afterwards, the spin algorithm continues as normal. We implement hard scattering
amplitudes for the processes considered in this work, i.e. $e^+ e^- \to \gamma^*
\to q \bar{q} g$, $H \to ggg$ and $H \to gq\bar{q}$.

The amplitudes are computed following the notation and conventions set out in
Ref.~\cite{Karlberg:2021kwr},
where a spinor product is denoted as 
\begin{equation}
  S_{\lambda}(p,q) = \bar{u}_{\lambda}(p) u_{-\lambda}(q).
\end{equation}
\subsection*{$e^+ e^- \to q \bar{q}g$} 
We denote the amplitude as $M(\lambda_{e^+}, \lambda_{e^-}; \lambda_{q}, \lambda_{\bar{q}}, \lambda_g)$, where the dependence
on the momenta is implicit. The non-vanishing helicity configurations are
\begin{align}
  M(\lambda,-\lambda;\lambda,-\lambda,\lambda) &= \frac{S_{-\lambda}(p_{e^-},p_{\bar{q}})^2 \, S_{\lambda}(p_{e^-}, p_{e^+})}{S_{-\lambda}(p_q,p_g) \, S_{-\lambda}(p_{\bar{q}},p_g)}, \\
  M(\lambda,-\lambda;\lambda,-\lambda,-\lambda) &= \frac{S_{\lambda}(p_{e^+},p_q)^2 \, S_{-\lambda}(p_{e^-}, p_{e^+})}{S_{\lambda}(p_q,p_g) \, S_{\lambda}(p_{\bar{q}},p_g)}, \\
  M(\lambda,-\lambda;-\lambda,\lambda,\lambda) &= -\frac{S_{-\lambda}(p_{e^-},p_q)^2 \, S_{\lambda}(p_{e^-}, p_{e^+})}{S_{-\lambda}(p_q,p_g) \, S_{-\lambda}(p_{\bar{q}},p_g)}, \\
  M(\lambda,-\lambda;-\lambda,\lambda,-\lambda) &= -\frac{S_{\lambda}(p_{e^+},p_{\bar{q}})^2 \, S_{-\lambda}(p_{e^-}, p_{e^+})}{S_{\lambda}(p_q,p_g) \, S_{\lambda}(p_{\bar{q}},p_g)}.
\end{align}

\subsection*{$H \to g g g$}
We denote the amplitude as $M(\lambda_{g_1}, \lambda_{g_2}, \lambda_{g_3})$. The non-vanishing helicity configurations are
\begin{align}
  M(\lambda,\lambda,\lambda) &= -\frac{1}{4} \frac{m_H^4}{S_{-\lambda}(p_{g_1}, p_{g_2}) S_{-\lambda}(p_{g_2}, p_{g_3}) S_{-\lambda}(p_{g_1}, p_{g_3})}, \\
  M(\lambda,\lambda,-\lambda) &= -\frac{S_{\lambda}(p_{g_1},p_{g_2})^3}{S_{\lambda}(p_{g_2}, p_{g_3})S_{\lambda}(p_{g_3}, p_{g_1})}, \\
  M(\lambda,-\lambda,\lambda) &= -\frac{S_{\lambda}(p_{g_3},p_{g_1})^3}{S_{\lambda}(p_{g_1}, p_{g_2})S_{\lambda}(p_{g_2}, p_{g_3})}, \\
  M(-\lambda,\lambda,\lambda) &= -\frac{S_{\lambda}(p_{g_2},p_{g_3})^3}{S_{\lambda}(p_{g_1}, p_{g_2})S_{\lambda}(p_{g_3}, p_{g_1})}.
\end{align}
\subsection*{$H \to g q \bar{q}$}
We denote the amplitude as $M(\lambda_{g}, \lambda_{q},
\lambda_{\bar{q}})$.
The non-vanishing helicity configurations are 
\begin{align}
  M(\lambda,\lambda,-\lambda) &= - \frac{S_{\lambda}(p_q,p_g)^2}{S_{\lambda}(p_q, p_{\bar{q}})}, \\
  M(\lambda,-\lambda,\lambda) &= \frac{S_{\lambda}(p_{\bar{q}},p_g)^2}{S_{\lambda}(p_q, p_{\bar{q}})}.
\end{align}
Note that this corresponds to the $H \to g q \bar{q}$ amplitude as
mediated by the heavy-top limit of the effective loop-induced $Hgg$
operator, as opposed to gluon emission following a direct $Hqq$ Yukawa
interaction.

\subsection{Validation}
\begin{figure}
  \centering
  \includegraphics[]{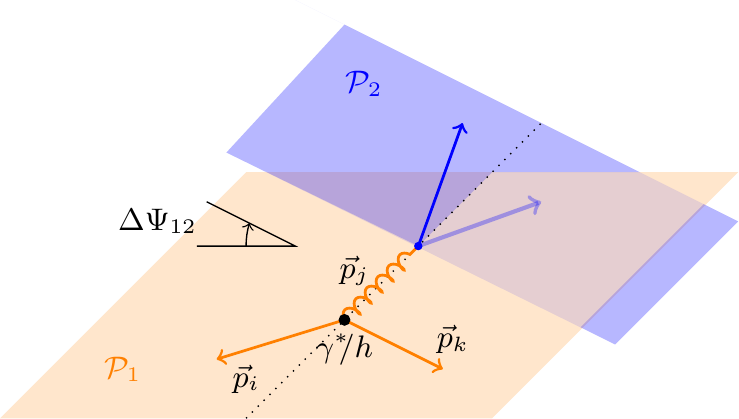}
  \caption{An illustration of the spin-sensitive observable $\Delta \Psi_{12}$
  as the difference of the azimuthal angles of the branching plane of the
  matched emission $\mathcal{P}_1$ and the branching plane of a subsequent
  collinear branching $\mathcal{P}_2$.}
  \label{fig:dpsi-definition}
\end{figure}
\begin{figure}
  \includegraphics[width=1.\textwidth]{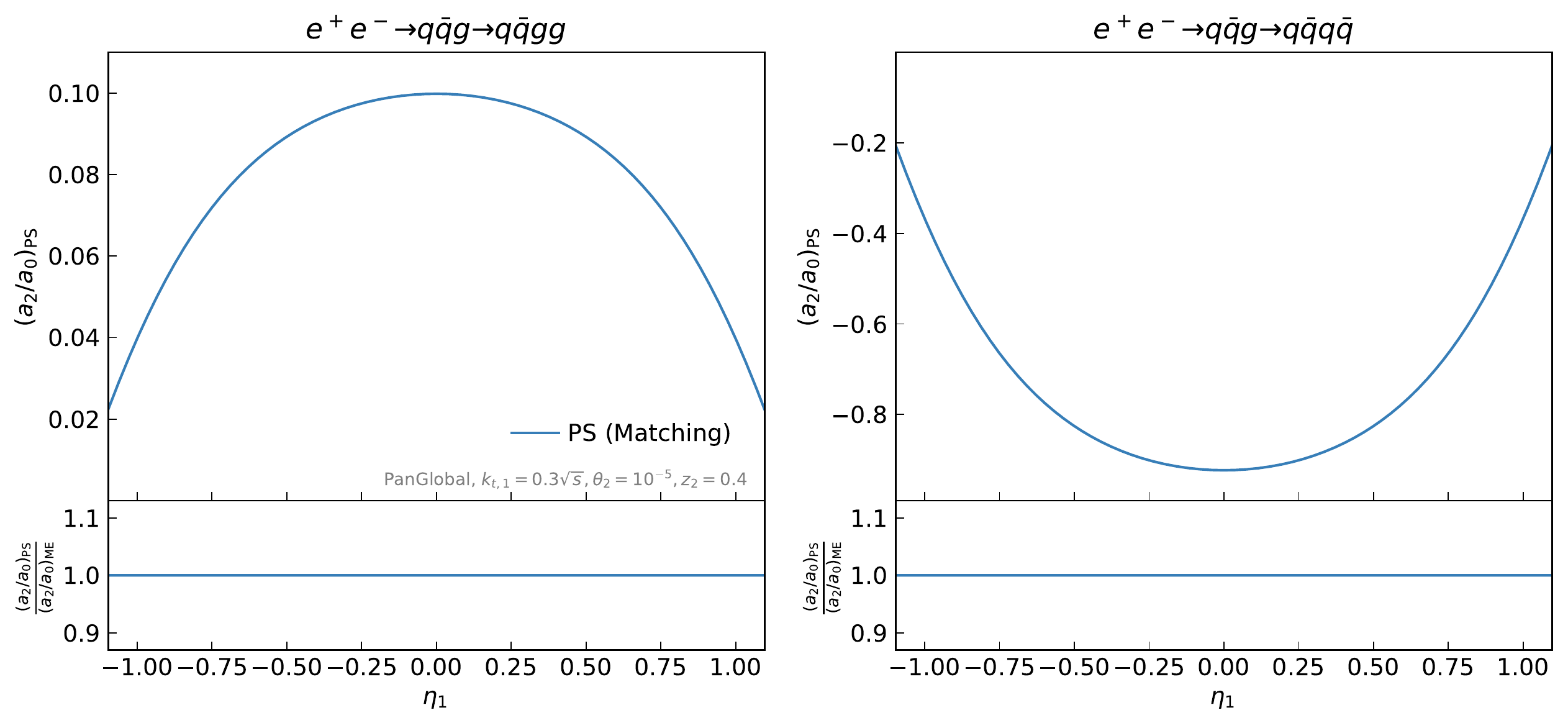}
  \caption{Second-order validation of the implementation of spin correlations in
  the hard region for $e^+ e^- \to \gamma^* \to q \bar{q}$. Following
  Eq.~\eqref{eq:hard-spin-xsec-form}, the ratio $a_2 / a_0$ is computed for the
  shower with matching enabled (blue curve). Below,
  the ratio of the shower $a_2 / a_0$ to the exact matrix element $a_2 / a_0$ is
  shown. }
  \label{fig:spin-qqbar}
\end{figure}
\begin{figure}
  \includegraphics[width=1.\textwidth]{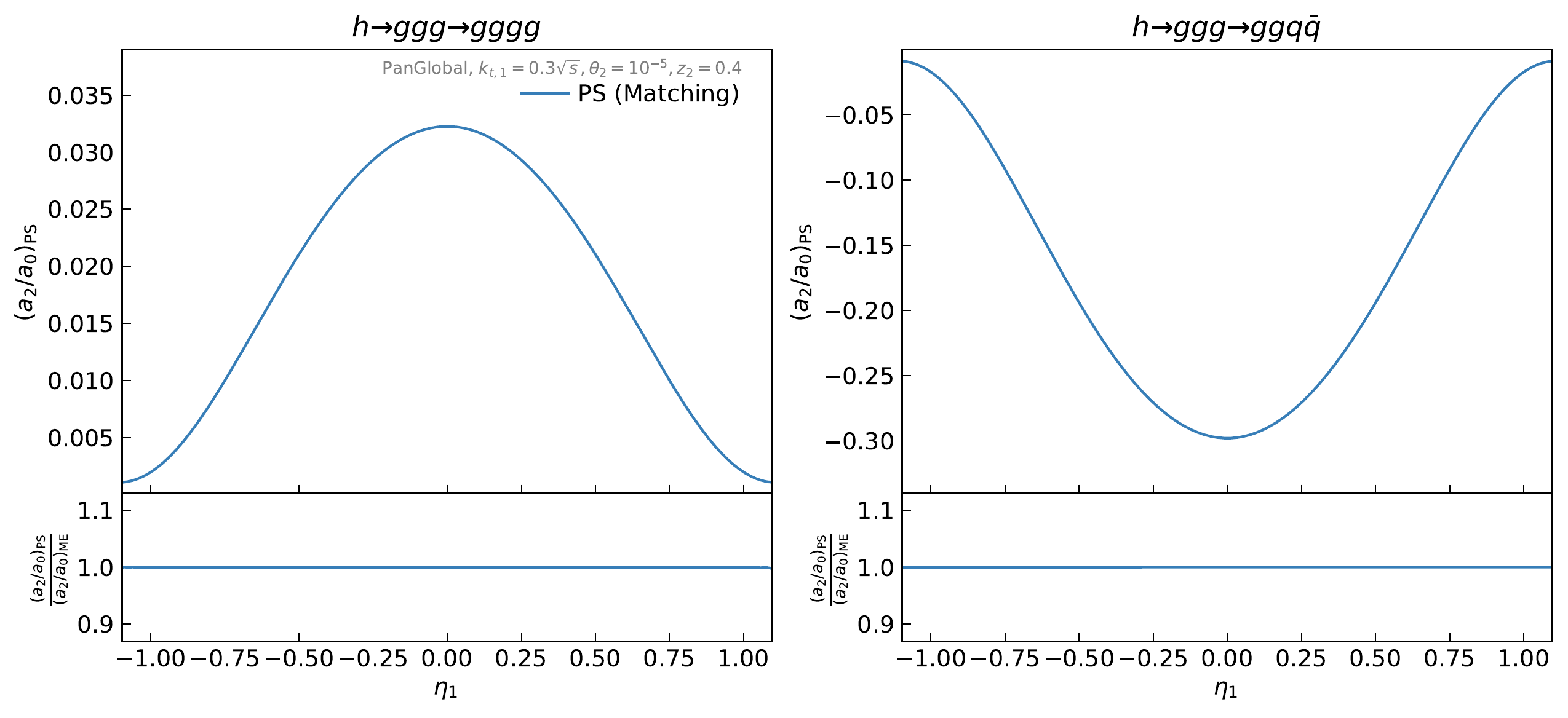}
  \includegraphics[width=1.\textwidth]{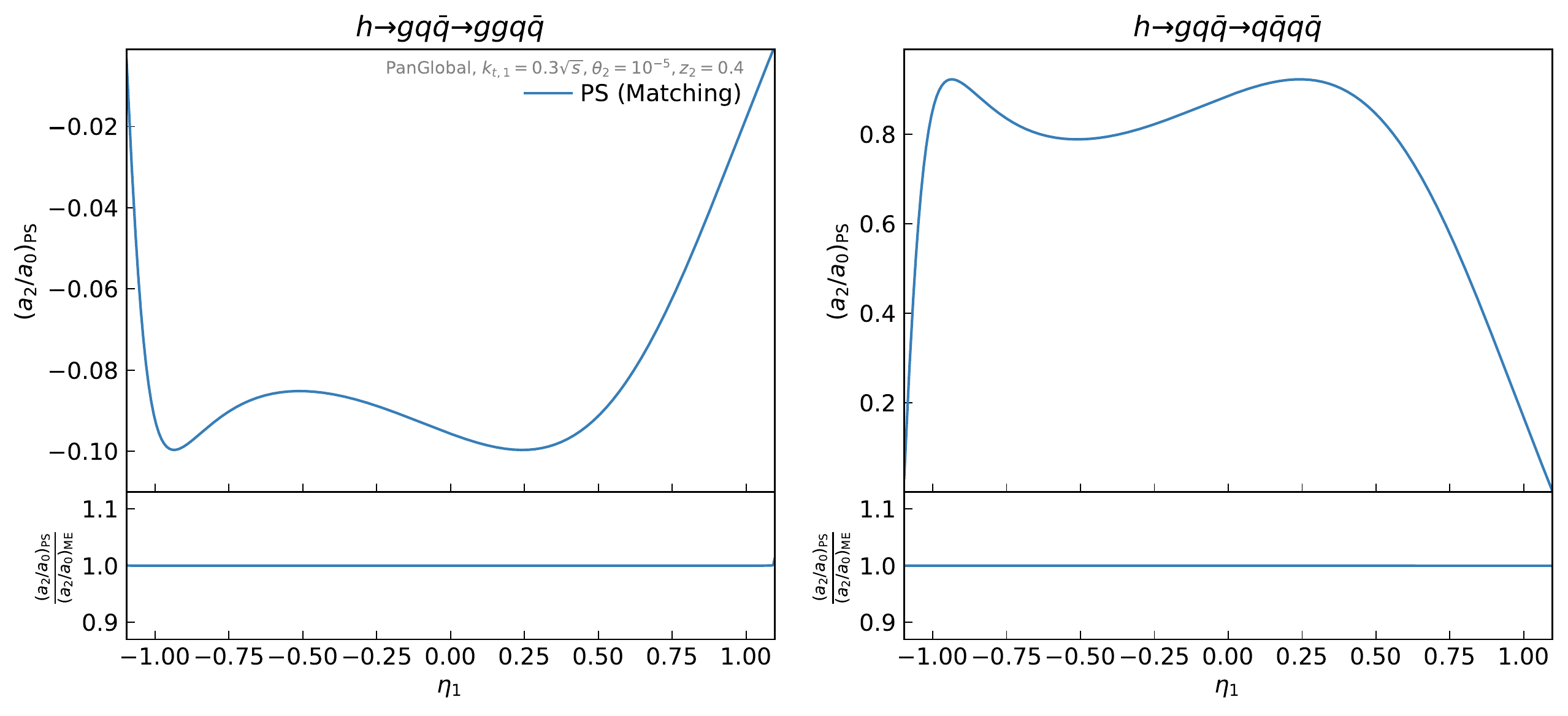}
  \caption{Same as Figure~\ref{fig:spin-qqbar}, but for all possible
    configurations of $H \to gg$.
    Note that in the lower plots, it is the gluon of the underlying
    3-jet system that undergoes collinear splitting, rather than the
    hard emission from the 2-jet system (which is a quark or
    anti-quark, and so does not mediate azimuthal correlation).
    This is also the reason for the lack of symmetry in the matched
    result between positive and negative $\eta_1$ values. 
  }
  \label{fig:spin-gg}
\end{figure}

We validate the implementation of the three-body hard scattering amplitudes at
second order by comparing to exact matrix elements of \texttt{MG5\_aMC@NLO}
\cite{Alwall:2014hca}. We consider configurations where the first matched
emission with momentum $p_k$ occurs from a dipole with momenta $\tilde{p}_i$ and
$\tilde{p}_j$ at a relatively large, fixed transverse momentum 
\begin{equation}
  k_{t,1}/Q =  \frac{\sqrt{p_i {\cdot} p_k \, p_j {\cdot} p_k}}{Q^2} = 0.3,
\end{equation}
where $p_i$ and $p_j$ are the post-branching dipole constituents, while varying
its rapidity
\begin{equation}
  \eta_1 = \frac{1}{2} \ln \frac{p_j{\cdot}p_k}{p_i {\cdot} p_k}.
\end{equation}
A second collinear branching then occurs with a small, fixed opening
angle $\theta_2 = 10^{-5}$ and collinear momentum fraction $z_2 = 0.4$.
When the $3$-jet event is produced via radiation of a gluon, it is
the collinear splitting of the radiated gluon that we examine; in $H\to gg$
events where the $3$-jet event is obtained through a $g \to q\bar q$
splitting, it is the collinear splitting of the remaining Born gluon
that we examine.
The difference between the azimuthal orientations of the
branching planes $\Delta \Psi_{12}$ is then sensitive to spin
correlations.
An illustration of this observable is shown in Fig.~\ref{fig:dpsi-definition}.
In particular, the cross section has the form 
\begin{equation} \label{eq:hard-spin-xsec-form}
  \frac{d \sigma}{d \Delta \Psi_{12} d\eta_1} \propto a_0(\eta_1) + a_2(\eta_1) \cos\left( 2 \Delta \Psi_{12} \right).
\end{equation}
The values of $a_0$ and $a_2$ can be extracted through a Fourier transform. The
results are shown in Fig.~\ref{fig:spin-qqbar} for $e^+ e^- \to \gamma^* \to q
\bar{q}$ and in Fig.~\ref{fig:spin-gg} for $H \to gg$. The ratio $(a_2 / a_0)$
is shown for the PanGlobal shower with matching enabled.
Furthermore, the ratio between the parton shower $(a_2 / a_0)$ and the exact
matrix element $(a_2 / a_0)$ is shown, confirming that the matched shower
reproduces the exact matrix element.

\section{Technical details of the matching implementation}
\label{sec:technical}

The tests performed in section~\ref{sec:log-tests} require us to probe very small
values of $\alpha_s$ and very large values of the relevant logarithm. In the
parton shower, this corresponds to very small values of the cutoff scale,
which in turn requires precise control over the numerical accuracy of the parton
shower. If we were to make use of the public implementations of \POWHEG
\cite{Alioli:2010xd} and \MCatNLO \cite{Alwall:2014hca}, we would lack this
precision. The results of section~\ref{sec:log-tests} are thus obtained with a
simple dedicated implementation of \POWHEG and \MCatNLO for the processes at
hand within the PanScales framework.

In the following, we provide some details concerning the implementation of
multiplicative matching in the PanScales showers, and of the implementation of \POWHEG
and \MCatNLO in the PanScales framework.

\subsection{Multiplicative matching}
\label{sec:internal-matching}

The multiplicative matching of the PanScales showers is accomplished by replacing the
usual splitting functions, which are only correct in the soft and collinear
limits, by $R(\Phi) / B_0(\PhiB)$ for the first emission only.
In situations where  the
shower partitions the total emission weight into multiple dipoles or antennae,
the same partitioning is applied to the weight $R(\Phi) /
B_0(\PhiB)$. 
Furthermore, a Jacobian
factor must be included to account for the transformation from the
radiative phase space to the shower variables.

For completeness, we give here the analytic expressions of the effective
one-emission matrix elements for the PanLocal dipole and PanGlobal showers,
as well as the Jacobian factor mentioned above. We also briefly discuss
a workaround for the issue of under-sampling in the PanGlobal shower,
which appears in hard configurations.

\subsubsection{PanLocal (dipole)}
\label{sec:technical-panlocal}

For the PanLocal dipole shower, the total radiation intensity is a sum over two dipole ends,
\begin{align}
  \frac{\Rps^\text{\sc pl}(\PhiBR)}{B_0(\PhiB)}
  \mathrm{d}\Phi_R
  &= {\rm d}\mathcal{P}_{\tilde i \tilde j}^{[\tilde i]} +
  {\rm d}\mathcal{P}_{\tilde i \tilde j}^{[\tilde j]}\nonumber\\
  &= {\rm d\,ln}v\, {\rm d}\bar \eta
     \frac{\alpha_s}{\pi} \left[ g(\bar\eta) a_k P_{\itilde \to ik}(a_k)
     + (\itilde \leftrightarrow \jtilde) \right] \nonumber \\
  &= {\rm d}x_i {\rm d}x_j \frac{\alpha_s}{\pi}
  \left[ \left| J^\text{\sc PL} \right|^{-1} g(\bar\eta) a_k P_{\itilde \to ik}(a_k)
     + (\itilde \leftrightarrow \jtilde) \right]\,,
\label{eq:panlocal_me}
\end{align}
where the dipole is partitioned at the bisector of the dipole's
opening angle (in the event frame) by a function $g(\bar\eta)$
(where $\bar\eta = \frac12 \ln \frac{a_k}{b_k}$),
\begin{equation}
  \bar \eta \leq -1: g(\bar \eta) \equiv 0,\quad
  -1 < \bar \eta < 1: g(\bar \eta) = \frac{15}{16} \left( \frac{\bar \eta^5}{5}
                      - \frac{2 \bar \eta^3}{3} +
                      \bar \eta + \frac{8}{15}\right),\,
  \quad \bar \eta \geq 1: g(\bar \eta) \equiv 1\,.
\end{equation}
The Dalitz variables $(x_i,x_j)$ are related to $(a_k,b_k)$ by
\begin{equation}
      x_i = 1-a_k + \frac{a_k b_k}{1-a_k}\,,
\quad x_j = 1-\frac{b_k}{1-a_k}\,.
\end{equation}
Finally, the Jacobian for the Panlocal dipole shower (for a single
dipole end) can be expressed as
\begin{equation}
  J^\text{\sc PL} = \left| \frac{\partial (x_i, x_j)}{\partial (\ln v, \bar
  \eta)} \right|
  = 2a_k b_k\frac{1-a_k-b_k}{(1-a_k)^2}
  = 2(1-x_i)(1-x_j)\,.
\label{eq:panlocal_jac}
\end{equation}

\subsubsection{PanGlobal}
\label{sec:technical-panglobal}

The one-emission matrix element for the PanGlobal shower is
similarly expressed as
\begin{align}
  \frac{\Rps^\text{\sc pg}(\PhiBR)}{B_0(\PhiB)}
  \mathrm{d}\Phi_R
  &= {\rm d}x_i {\rm d}x_j \frac{\alpha_s}{\pi}
  \left[ \left| J^\text{\sc PG} \right|^{-1} f(\bar\eta) a_k P_{\itilde \to ik}(a_k)
     + (\itilde \leftrightarrow \jtilde) \right]\,,
\label{eq:panglobal_me}
\end{align}
with a partition function $f(\bar\eta)$ for the fraction of transverse
momentum recoil shared across two elements of the antenna,
\begin{equation}
f = f(\bar{\eta}) = \frac{e^{2\bar{\eta}}}{1+e^{2\bar{\eta}}}\,.
\end{equation}
Here the relation between the Dalitz and the shower variables is given by
\begin{equation}
      x_i = \frac{1-a_k}{1-a_k b_k}\,,
\quad x_j = \frac{1-b_k}{1-a_k b_k}\,,
\end{equation}
and the Jacobian for PanGlobal can be shown to be
\begin{equation}
J^\text{\sc PG} = \left| \frac{\partial (x_i, x_j)}{\partial (\ln v,
\bar \eta)} \right| = \frac{2a_k b_k (1-a_k)(1-b_k)}{(1-a_kb_k)^3} = \frac{2 x_i
x_j (1-x_i) (1-x_j)}{x_i + x_j - 1}\,.
\label{eq:panglobal_jac}
\end{equation}
The Jacobian has a divergence on the contour $x_i + x_j = 1$, or
equivalently at the point $\ln v = 0$. In other words, the
inverse Jacobian vanishes on that same contour, and the shower
has an emission probability that is exactly equal to zero on
that line. We tackle this issue by replacing the shower's
sampling of $\ln v$ values:
\begin{equation}
{\rm d\,ln}v\, {\rm d}\bar
\eta \frac{\alpha_s}{\pi} \rightarrow {\rm d\,ln}v\, {\rm d}\bar \eta
\frac{\alpha_s}{\pi} \frac{|\ln \frac{v}{Q}| + \mathcal{C}}{|\ln \frac{v}{Q}|}\,,
\end{equation}
with $\mathcal{C} > 0$. This modification only impacts the hard region,
$(|\ln \frac{v}{Q}| + \mathcal{C})/|\ln \frac{v}{Q}| \to 1$ when $\ln \frac{v}{Q} \to -\infty$. 
On the problematic contour, the modified radiation intensity becomes
\begin{equation}
\lim_{x_i+x_j\to 1} {\rm d}\mathcal{P}_{\tilde i \tilde j}^{[\tilde i]} = {\rm
d}x_i {\rm d}x_j \frac{\alpha_s C}{2\pi} \frac{\mathcal{C}}{x_i x_j
(1+\betaps|x_i-x_j|)}\,\,
\end{equation}
and is thus non-zero for $\mathcal{C} > 0$. In practice we take $\mathcal{C} = \frac12$.

\subsection{\MCatNLO}
\label{sec:MCatNLO-appendix}
The implementation of \MCatNLO amounts to the production of a mixture
of two sets of events, one distributed according to the Born matrix
element, the other according to the real correction term.
Rather than implementing a separate phase space generator for the second set, a
simpler method is possible when, following
section~\ref{sec:internal-matching} a multiplicatively-matched version of the shower
is already available. In that case, we can use the shower as the phase space
generator. Specifically, we employ a first-order expansion of the shower, in
which a radiating dipole is selected randomly with equal probability.
Then, event weights equal to the
difference between the matched and unmatched weights of the radiating dipoles
are attached to the resulting event.

Our normalisation convention differs slightly from the standard 
\MCatNLO approach, mainly for reasons of programmatic convenience.
Specifically we use 
\begin{multline}
  \label{eq:MCatNLO-altnorm}
  \mathd \sigma_\text{\MCatNLO} =
  \frac{\BB}{B_0 + (\BB - \BB_\ps)} \Bigg[
       S_\ps(v_\Phi^\ps,\PhiB)
      \times {\Rps(\PhiBR)}  \, \mathd \PhiBR \times I_\ps(v_\Phi^\ps, \PhiBR)
                \,+\\+ \left[R(\PhiBR)-\Rps{}(\PhiBR)\right] \mathd
                \PhiBR \times I_\ps(v^{\max}, \PhiBR) 
                \Bigg],
\end{multline}
where $B_0$, $\BB$ and $\BB_\ps$ are all to be understood as being
functions of $\Phi_B$.
It is straightforward to verify that this differs from
Eq.~(\ref{eq:MCatNLO}) only starting from order $\as^2$ relative to
the Born cross section, i.e.\ NNLO.

\subsection{PanScales adaptation of FKS map (\powhegbeta)}
\label{sec:appPOWHEG}

In the PanScales framework we implement the FKS map, broadly similar
to that used in the {\tt
  POWHEG-BOX} by casting it into the form of a shower that handles the
first emission only. The kinematic map can be written as
\begin{subequations}\label{eq:powheg-map}
  \begin{align} 
    p_i &= a_i \tilde{p}_i + b_i \tilde{p}_j + k_{\perp}\,,  \\
    p_j &= b_j \tilde{p}_j\,,  \\
    p_k &= a_k \tilde{p}_i + b_k \tilde{p}_j - k_{\perp}\,.
  \end{align}
\end{subequations}
The FKS map parameterises the radiative phase space in terms of 
\begin{equation} \label{eq:powheg-variables}
\xi = 2\frac{p_k^0}{\sqrt{s}},  \quad y = \cos \theta_{ik}\,,
\end{equation}
and an azimuthal angle $\varphi$ that determines the orientation of
the transverse momentum $k_{\perp}$. The coefficients of the map of
Eq.~\eqref{eq:powheg-map} are then completely fixed by requiring the momenta to
adhere to Eq.~\eqref{eq:powheg-variables}, $p_i^2 = p_j^2 = p_k^2 = 0$ and
momentum conservation.
Furthermore, our \POWHEGbeta map supports a more general ordering
variable, designed to coincide with the PanScales maps in both the
soft-collinear and the soft large-angle regions.
To that end the phase space
is reparameterised in terms of 
\begin{equation} \label{eq:powheg-panscales-variables}
  \bar \eta = - {\rm ln\, tan} \left( \frac{{\rm arccos\,}y}{2}
\right)\,,\quad\ln v = \ln \frac{\sqrt{s}}{2} + \ln \sin \big[2 \arctan e^{-\bar
\eta} \big] +  \ln \xi - \beta |\bar \eta|\,.
\end{equation}
Emissions are then generated as in a normal shower, but with a total splitting
weight $R(\Phi) / B_0(\PhiB)$, where we also include the appropriate Jacobian
associated with the transformation from the radiative phase space to the
parameterisation of Eq.~\eqref{eq:powheg-panscales-variables}.
The splitting
weight is partitioned into terms associated with pairs of partons becoming
collinear to one another, according to the FKS
prescription~\cite{Frixione:2007vw,Frixione:1997np}, and the contribution of
external gluons is partitioned into two dipoles following the discussion in
section \ref{sec:nndl-requirements}.
The two dipoles are allowed to radiate
independently, ensuring the total weight reproduces $R(\Phi) / B_0(\PhiB)$.
%

\bibliographystyle{JHEP}
\bibliography{MC}

\end{document}